\documentclass[12pt,a4paper]{article}

\usepackage{latexsym}
\usepackage[dvips]{graphics}
\usepackage{epsfig}
\usepackage{color}
\usepackage{amsmath,amssymb}
\usepackage{cite,mcite}
\usepackage{xspace}
\usepackage{ifthen}
\usepackage{hyperref}
\usepackage{slashed}
\usepackage{epstopdf}
\usepackage[rightcaption]{sidecap}
\sidecaptionvpos{figure}{c}

\usepackage[hang,small]{caption}
\usepackage{here}
\usepackage{placeins}

\usepackage{tabularx}

\newcommand{\beq}{\begin{equation}}
\newcommand{\eeq}{\end{equation}}
\newcommand{\bea}{\begin{eqnarray}}
\newcommand{\eea}{\end{eqnarray}}

\newcommand{\be}{\begin{eqnarray*}}
\newcommand{\ee}{\end{eqnarray*}}

\newcommand{\spin}[1]{\mbox{spin-#1}}

\newcolumntype{C}[1]{>{\centering\arraybackslash}p{#1}}

\numberwithin{equation}{section}

\begin{document}

\thispagestyle{empty}

\begin{center}
\hfill KA-TP-40-2012 \\
\hfill SFB/CPP-12-86 
\end{center}

\begin{center}

\vspace{1.7cm}

{\LARGE\bf 
Spin-2 Resonances in Vector-Boson-Fusion Processes at NLO QCD} 

\vspace{1.4cm}

{\bf Jessica Frank, 
\bf Michael Rauch, 
\bf Dieter Zeppenfeld}\\ 

\vspace{1.2cm}

{\em {Institute for Theoretical Physics, Karlsruhe 
Institute of Technology, 76128 Karlsruhe, Germany}
}\\

\end{center}

\vfill

\centerline{\bf Abstract}
\vspace{2 mm}
\begin{quote}
\small
The most likely spin assignments of the recently discovered 126 GeV resonance
are spin 0 or 2. In order to distinguish the two, we construct an effective
Lagrangian model which comprises interactions of a \spin2 electroweak
singlet or triplet state with the SM gauge bosons. Within
this model, cross sections and differential distributions are calculated 
and implemented within the Monte
Carlo program \textsc{Vbfnlo}, which simulates vector-boson-fusion
processes at hadron colliders at NLO QCD accuracy.
We study the phenomenology of \spin2 resonances produced in
vector-boson-fusion processes at the LHC.  Specifically, we consider light
Higgs-like \spin2 resonances decaying into two photons and show how angular 
distributions allow us to distinguish between a Standard Model Higgs and
a \spin2 resonance.  We also investigate the characteristics of 
heavy \spin2 resonances which decay into two weak gauge bosons, leading to a
four-lepton final state.  
\end{quote}

\vfill

\newpage
%
\section{Introduction \label{sec:Intro}}
One of the main tasks of the LHC is to uncover the origin of electroweak
symmetry breaking. A giant step in this direction was made recently with
the observation of a narrow resonance decaying into pairs of electroweak
gauge bosons at an invariant mass of about 126~GeV by the LHC
experiments ATLAS~\cite{ATL-CONF-comb} and CMS~\cite{CMS-CONF-comb}.
The data obtained for this resonance, as yet, are compatible with it
being a SM Higgs boson~\cite{Higgs}. One of the main tasks for the
coming years is to perform increasingly stringent tests of the assertion
that, indeed, the SM Higgs boson has been found.

There are several possible channels for the detection of the Higgs resonance
and for measurements of its
properties.  A production channel with fairly large cross section
is vector-boson-fusion (VBF)~\cite{vbf}, which shows a clear signature of
two highly energetic jets in the forward and backward regions of the detector.
The loop-induced decay to a final state with two photons then
allows a full reconstruction of the Higgs' four-momentum, leading to a
sharp peak in the di-photon mass spectrum. Thereby, it is possible to
distinguish the signal from the large background despite its small
branching ratio, and a $\gamma\gamma$ peak has already been seen in 
the VBF analysis of CMS~\cite{cmsvbfaa}.

For a definite verification that it is indeed the SM Higgs boson which 
has been discovered at CERN, all features of the
Higgs boson need to be tested~\cite{Burgess:1999ha}. These are its
couplings~\cite{SFitterHiggs} (including its
self-couplings~\cite{Higgsself}), its CP quantum number and its
spin~\cite{HiggsCP}.
Observation of this resonance in the di-photon channel immediately
excludes a \spin1 particle due to the Landau-Yang
theorem~\cite{Landau-Yang}. Besides the \spin0 of the Higgs boson, 
a \spin2 particle would also be possible. Since the distinction of a \spin0 and
a \spin2 resonance is an important task for Higgs physics, this paper provides
tools for differentiating between the two. Specifically, we use an 
effective Lagrangian for a \spin2 field interacting with electroweak 
gauge bosons to calculate VBF cross sections and 
distributions at NLO QCD precision. These calculations are implemented
in the \textsc{Vbfnlo} program~\cite{Arnold:2011wj}, which is then used to search
for characteristic differential distributions which distinguish between
the two spin choices.  Our analysis focuses on angular correlations,
since they are known as a powerful tool to study the spin of a
resonance. 

Due to the high energies which can be achieved with the LHC, it might
also be possible to detect some new, heavy resonances in VBF processes,
which are manifestations of physics beyond the Standard Model. For these
resonances, a spin determination would also be needed. Whereas heavy
\spin1 resonances have already been studied within the \textsc{Vbfnlo}
framework~\cite{Englert:2008}, our present analysis will consider the
characteristics of heavy \spin2 resonances. 

Within the present analysis, the features of \spin2 resonances in
VBF are studied for two cases: light, Higgs-like
resonances in the photon pair-production channel and heavy resonances in
processes with four leptons in the final state 
(in addition to the two tagging jets characterizing VBF). 
These are
$e^+ \, e^- \, \mu^+ \mu^- \, jj$, $\, \, e^+ \, e^- \, \nu_\mu
\overline{\nu}_\mu \, jj$, $\, \, e^+ \, \nu_e \, \mu^- \,
\overline{\nu}_\mu \, jj$, $ \, \, e^+ \, \nu_e \, \mu^+ \mu^- \, jj$
and $e^- \, \overline{\nu}_e \, \mu^+ \mu^- \, jj$. Of these, the first
one will be studied in most detail, since a final state which does not
contain neutrinos allows for a full reconstruction of a resonance. In
order to describe the interaction of \spin2 particles with electroweak
gauge bosons, we have constructed an effective model which comprises two
different scenarios: a \spin2 state which behaves as a singlet under
SU(2) transformations and a \spin2 triplet. 

This paper is organized as follows: 
In Section~\ref{sec:spin2model}, we present our model for the
interaction of neutral and charged \spin2 particles with electroweak
bosons. The relevant aspects of our calculation, including our choice of
input parameters and selection cuts, are sketched in
Section~\ref{sec:Calculation}. 
In Section~\ref{sec:Results} the results of our analysis are presented,
where we show the characteristics of \spin2 resonances in the different
kinds of processes.
Additionally, we discuss to what extent they can be used for a
distinction of a \spin2 resonance from a Higgs boson or the SM
non-resonant background, respectively.  Furthermore, the impact 
of the NLO QCD corrections is examined.  
Conclusions are drawn in Section~\ref{sec:conclusions}. 
Specific formulas describing the decay widths of the \spin2 particles
are given in the Appendix.

 \section{\label{sec:spin2model} The Spin-2 Model}

For the present analysis of \spin2 resonances in vector-boson-fusion
processes, we have constructed an effective model describing the
interaction of \spin2 particles with electroweak bosons.  Two cases 
are considered: A \spin2 state which behaves as a singlet under SU(2)
transformations  and a \spin2 state which is a weak isospin triplet.

These states are described by the general \spin2 fields $T^{\mu \nu}$
(singlet) and $T^{\mu \nu}_j$ (triplet),
\begin{equation}
T_{(j)}^{\mu \nu}(x)=\int{\frac{d^3k}{(2\pi)^3 2k^0}}\sum_{\lambda=-2}^2{\left( \varepsilon^{\mu \nu}(k,\lambda)a_{\lambda (,j)}(k)e^{-ikx}+
\varepsilon^{* \mu \nu} (k, \lambda) a_{\lambda (,j)}^\dagger (k) e^{ikx} \right)}. \label{spin2 field}
\end{equation}
The free Lagrangian for a general \spin2 field with mass $m$ is given by~\cite{freespin2}
\begin{equation}
\mathcal{L}_\text{free} = 
- \left(\partial_\mu T^{\mu\nu} \right)^\dagger \left(\partial_\rho {T^{\rho}}_{\nu} \right) 
+ \frac12 \left(\partial_\rho T^{\mu\nu} \right)^\dagger \left(\partial^\rho T_{\mu\nu} \right)
+ \frac{m^2}2 \; {T^{\mu\nu}}^\dagger T_{\mu\nu} \; .
\end{equation}
For the triplet field, the partial derivatives are to be replaced by
covariant ones in order to account for its gauge couplings to
electroweak bosons. Note, however, that these couplings induce $TTV$ or
$TTVV$ vertices, which do not appear in the processes studied in this
paper.
The fields are symmetric in $\mu, \nu$, transverse and $T_\mu^{\mu}=T_\mu^{\mu, j}=0$.
$\varepsilon^{\mu \nu}$ is a symmetric polarization tensor built from
the usual \spin1 polarization vectors~\cite{Hagiwara:2008jb}:
\begin{align}
\epsilon^{\mu\nu}(p,\pm2) &= \epsilon^{\mu}(p,\pm) \epsilon^{\nu}(p,\pm)
\nonumber\\
\epsilon^{\mu\nu}(p,\pm1) &= \frac1{\sqrt{2}} \left(
\epsilon^{\mu}(p,\pm) \epsilon^{\nu}(p,0) +
\epsilon^{\mu}(p,0) \epsilon^{\nu}(p,\pm) \right) \nonumber\\
\epsilon^{\mu\nu}(p,0) &= \frac1{\sqrt{6}} \left(
\epsilon^{\mu}(p,+) \epsilon^{\nu}(p,-) +
\epsilon^{\mu}(p,-) \epsilon^{\nu}(p,+) +
2 \epsilon^{\mu}(p,0) \epsilon^{\nu}(p,0) \right) \ .
\end{align}

While the \spin2 singlet involves only one uncharged particle, called $T$, the
triplet consists of three \spin2 particles, $T^1$, $T^2$ and $T^3$, or,
equivalently,  a charged pair and a neutral particle:
\begin{align}
T^\pm &=\frac{1}{\sqrt{2}} (T^1 \mp i \,T^2), \nonumber\\[0.1cm]
T^0 &=T^3.
\end{align}

Since in the present analysis we only study spin-2 resonances which 
are produced in electroweak-boson fusion, we restrict ourselves to 
a model of the interaction of a single \spin2 particle with electroweak
bosons. 
The building blocks of the corresponding singlet and triplet Lagrangian
were chosen to be the \spin2 field(s), the vector fields of the
electroweak gauge bosons and the scalar Higgs field $\Phi$. 
Respecting gauge and Lorentz invariance 
and neglecting higher dimensional operators, we end up with the
following effective Lagrangian for the singlet case:

\begin{align}
 \mathcal{L}_{\text{singlet}}&=\frac{1}{\Lambda} T_{\mu \nu} \left(f_1 B^{\alpha \nu} B^\mu_{\hspace{0.15 cm} \alpha} +f_2 W_i^{\alpha \nu} W^{i, \mu}_{\hspace{0.3 cm} \alpha}+ 2f_5 (D^\mu\Phi)^\dagger(D^\nu\Phi) \right), \label{spin2lagrangian}
\end{align}
while the Lagrangian corresponding to the triplet case reads
\begin{align}
\mathcal{L}_{\text{triplet}}&=\frac{1}{\Lambda} T_{\mu \nu, j} \left(f_6 (D^\mu\Phi)^\dagger \sigma^j (D^\nu\Phi) +f_7  W^{j, \mu}_{\hspace{0.3 cm} \alpha} B^{\alpha \nu}
 \right) \label{tripletlagrangian}.
\end{align}
${\Lambda}$ is the characteristic energy scale of the underlying new
physics, $f_i$ are variable coupling parameters, $B^{\alpha \nu}$ and
$W_i^{\alpha \nu}$ are the usual electroweak field strength tensors and
$D^\mu$ is the covariant derivative

\begin{equation}
D^\mu=\partial^\mu-igW_i^\mu \frac{\sigma^i}{2} -ig'Y B^\mu.
\end{equation}
The terms in Eq.~(\ref{spin2lagrangian}) and~(\ref{tripletlagrangian}) exhaust the possible parity-conserving 
contributions at the dimension five level.
The masses of the \spin2 particles are taken as free parameters.

In contrast to the graviton Lagrangian \cite{Hagiwara:2008jb}, couplings
to fermions or gluons are not included in the Lagrangians
(\ref{spin2lagrangian}, \ref{tripletlagrangian}).
Another important difference to the graviton Lagrangian is the presence
of variable prefactors $f_i$, which are not fixed to the same value by
the underlying theory.

It is possible to write down additional terms including the dual 
electroweak field strength tensors $\widetilde{B}^{\alpha \nu} = 
\frac{1}{2} \varepsilon^{\alpha \nu \rho \sigma} B_{\rho \sigma}$ and
$\widetilde{W}_i^{\alpha \nu}$, namely $\frac{f_3}{\Lambda} T_{\mu \nu}
\widetilde{B}^{\alpha \nu} B^\mu_{\hspace{0.15 cm} \alpha}$ and
$\frac{f_4}{\Lambda} T_{\mu \nu} \widetilde{W}_i^{\alpha \nu} W^{i
\mu}_{\hspace{0.3 cm} \alpha}$. However, such terms yield $TVV$ vertices 
which are proportional to $T^\mu_\mu$ and, thus, vanish for on-shell 
\spin2 particles. Off-shell
contributions do not lead to significant observable effects in the
distributions to be studied below.

The four relevant vertices resulting from the singlet Lagrangian
(\ref{spin2lagrangian}), which involve two electroweak bosons and the
\spin2 particle $T$, are $TW^+W^-$, $TZZ$, $T\gamma \gamma$ and $T\gamma
Z$. The calculation of the corresponding Feynman rules yields the 
following expressions for the $TVV$ vertices:

\begin{align}
TW^+W^- &:\,\, \frac{2i f_2}{\Lambda} K_1^{\alpha \beta \mu \nu} +\frac{if_5 g^2 v^2}{2 \Lambda} K_2^{\alpha \beta \mu \nu}, \nonumber\\
TZZ &:\,\, \frac{2i}{\Lambda} (f_2 c_w^2+f_1 s_w^2) K_1^{\alpha \beta \mu \nu} + \frac{if_5 v^2}{2 \Lambda} (g^2+g'^2) K_2^{\alpha \beta \mu \nu}, \nonumber\\
T\gamma \gamma &:\,\, \frac{2i}{\Lambda} (f_1 c_w^2+f_2 s_w^2) K_1^{\alpha \beta \mu \nu}, \nonumber\\
T\gamma Z &:\,\, \frac{2i}{\Lambda} c_w s_w (f_2-f_1) K_1^{\alpha \beta \mu \nu},
\label{Feynmanrules}
\end{align}
where $c_w$ and $s_w$ denote the cosine and sine of the Weinberg angle,
$v$ is the vacuum expectation value of the Higgs field and the two different tensor structures are given by
\begin{align}
K_1^{\alpha \beta \mu \nu}&=p_1^\nu \, p_2^\mu \, g^{\alpha \beta}- p_1^\beta \, p_2^\nu \, g^{\alpha \mu}-
p_2^\alpha \, p_1^\nu \, g^{\beta \mu}+p_1 \cdot p_2 \, g^{\alpha \nu} g^{\beta \mu},\nonumber\\
K_2^{\alpha \beta \mu \nu}&=g^{\alpha \nu} g^{\beta \mu}.
\label{tensorstructures}
\end{align}
The indices $\mu$ and $\nu$ correspond to the \spin2 field (which is
symmetric), $\alpha$ is the index of the first electroweak boson, whose
incoming four-momentum is denoted as $p_1$, and $\beta$ is the index of
the second one with four-momentum $p_2$.

The triplet Lagrangian (\ref{tripletlagrangian}) yields the same
relevant vertices for the uncharged \spin2 particle $T^0$ as the singlet
Lagrangian and two additional relevant vertices for the charged
particles $T^+$ and $T^-$. The structure of the Feynman rules is
analogous to the singlet case:

\begin{align}
T^0 W^+W^- &:\,\, \frac{i f_6}{4 \Lambda}g^2 v^2 K_2^{\alpha \beta \mu \nu}, \nonumber\\
T^0 ZZ &:\,\, -\frac{if_6}{4 \Lambda} (g^2+g'^2) v^2 K_2^{\alpha \beta \mu \nu} -\frac{2i f_7}{\Lambda} c_w s_w K_1^{\alpha \beta \mu \nu}, \nonumber\\
T^0 \gamma \gamma &:\,\, \frac{2i f_7}{\Lambda} c_w s_w K_1^{\alpha \beta \mu \nu}, \nonumber\\
T^0 \gamma Z &:\,\, \frac{i f_7}{\Lambda} (c_w^2 - s_w^2) K_1^{\alpha \beta \mu \nu}, \nonumber
\end{align}
\begin{align}
T^\pm W^\mp Z &:\,\, -\frac{if_6}{4 \Lambda}g v^2 \sqrt{g^2+g'^2} K_2^{\alpha \beta \mu \nu} -\frac{i f_7}{\Lambda} s_w K_1^{\alpha \beta \mu \nu}, \nonumber\\
T^\pm W^\mp \gamma &:\,\, \frac{i f_7}{\Lambda} c_w K_1^{\alpha \beta \mu \nu},
\label{tripletFeynmanrules}
\end{align}
with $K_1^{\alpha \beta \mu \nu}$ and $K_2^{\alpha \beta \mu \nu}$
defined as in the singlet case (Eq. (\ref{tensorstructures})).

The propagator of the \spin2
field with momentum $k$, i.e.\ the Fourier transform of 
$\left<0\left|\mathcal{T}\left(T^{\mu\nu}(x)T^{\alpha\beta}(0)\right)\right|0\right>$, 
is given by~\cite{Hagiwara:2008jb,Giudice:1998ck}
\begin{equation}
\frac{i B^{\mu \nu\alpha\beta}(k)}{k^2-m_T^2+im_T\Gamma_T}, \label{propagator1}
\end{equation}\\
where $m_T$ is the mass of the \spin2 particle, $\Gamma_T$ is its width 
and $B^{\mu \nu\alpha\beta}(k)$ is defined as
\begin{align}
B^{\mu \nu\alpha\beta}(k)&=\frac{1}{2}\left(g^{\mu\alpha}g^{\nu\beta}+g^{\mu\beta}g^{\nu\alpha}-g^{\mu\nu}g^{\alpha\beta}\right)
+\frac{1}{6}\left(g^{\mu\nu}+\frac{2}{m_T^2}k^\mu k^\nu\right)\left(g^{\alpha\beta}+\frac{2}{m_T^2}k^\alpha k^\beta \right)\nonumber\\
&\quad -\frac{1}{2m_T^2}\left(g^{\mu\alpha}k^\nu k^\beta+g^{\nu\beta}k^\mu k^\alpha+g^{\mu\beta}k^\nu k^\alpha+g^{\nu\alpha}k^\mu k^\beta\right).
\label{propagator2}
\end{align}
Explicit expressions for the partial decay widths into all possible
final states are given in App.~\ref{sec:decaywidths}.

Since the present \spin2 model is based on an effective Lagrangian
approach, it violates unitarity above a certain energy scale. In order
to parametrize high-energy contributions beyond this effective model, we
introduce the following formfactor, which can be multiplied with the
amplitudes:
\begin{equation}
f(p_1^2, p_2^2, k_{\text{sp2}}^2)=\left( \frac{\Lambda_{ff}^2}{|p_1^2|+\Lambda_{ff}^2} \cdot \frac{\Lambda_{ff}^2}{|p_2^2|+\Lambda_{ff}^2} \cdot 
\frac{\Lambda_{ff}^2}{|k_{\text{sp2}}^2|+\Lambda_{ff}^2} \right)^{n_{ff}}, \label{formfactor}
\end{equation}
where $p_1^2$ and $p_2^2$ are the invariant masses of the initial
electroweak bosons and $k_{\text{sp2}}^2$ is the squared invariant mass of the
sum of the initial boson momenta, equivalent to that of an $s$-channel
\spin2 particle. The energy scale $\Lambda_{ff}$ and the exponent
$n_{ff}$ are free parameters, describing the scale of the
cutoff and the suppression power, respectively.

\FloatBarrier\section{\label{sec:Calculation} {Elements of the Calculation}}

For the present analysis, we use the parton-level Monte Carlo program
\textsc{Vbfnlo} \cite{Arnold:2011wj}, which simulates
vector-boson-fusion processes at hadron colliders with NLO QCD accuracy. 
The characteristics of \spin2 resonances are studied for two different classes of processes: 
VBF photon pair-production in association with two jets and processes
with four leptons and two jets in the final states, namely $e^+ \, e^-
\, \mu^+ \mu^- \, jj$, $\, \, e^+ \, e^- \, \nu_\mu \overline{\nu}_\mu
\, jj$, $\, \, e^+ \, \nu_e \, \mu^- \, \overline{\nu}_\mu \, jj$, $ \,
\, e^+ \, \nu_e \, \mu^+ \mu^- \, jj$ and $e^- \, \overline{\nu}_e \,
\mu^+ \mu^- \, jj$ production.

The VBF processes with four leptons and two jets in the final state have
already been analyzed at NLO QCD accuracy within the SM. These
calculations, which are described in Refs.\ \cite{Jager:2006zc},
\cite{Jager:2006cp} and \cite{Bozzi:2007ur}, have been extended by the
effects of the \spin2 model for the present analysis.    
Both resonant and non-resonant sub-processes in typical VBF phase-space regions are considered. 
The contributing Feynman graphs at tree-level can be grouped into different topologies, 
where either one, two or three electroweak bosons are attached to the same quark line. Quark--anti-quark initiated $t$-channel processes, resulting from 
crossing the respective quark-quark diagrams, and $u$-channel diagrams obtained by interchanging identical initial- or final-state quarks,
 are also fully taken into account. However, interference between $t$- and $u$-channel contributions can safely be neglected in VBF phase-space regions.
 $s$-channel exchange, which corresponds to triple vector-boson production, with one of the time-like bosons decaying into a pair of jets, is considered as 
a separate process in \textsc{Vbfnlo} which, however, is strongly suppressed in VBF phase-space regions and will not be considered in the following. 

\begin{SCfigure}[\sidecaptionrelwidth][ht]
\includegraphics[width=0.53\textwidth]{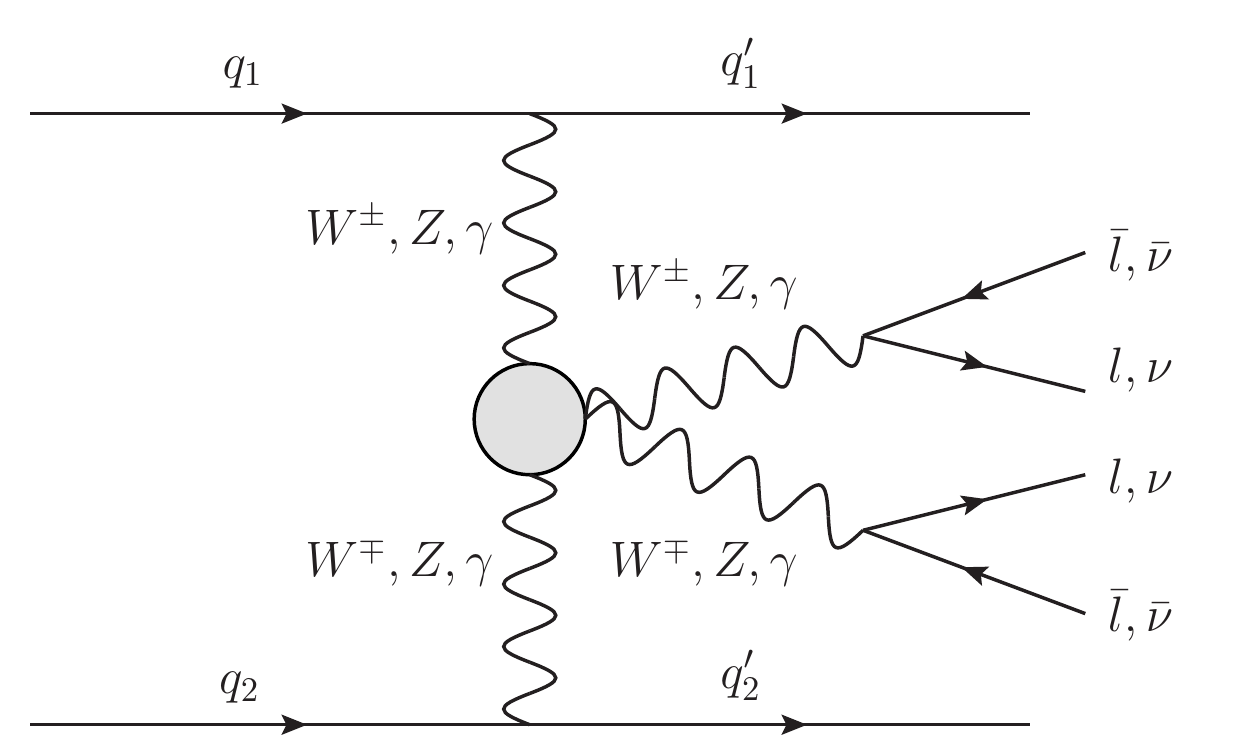}
\caption{General vector-boson-fusion Feynman graph at tree-level, where
\spin2 effects can appear.}
\label{vbffig}
\end{SCfigure}

The only tree-level topology in which the \spin2 particles of our model 
can be exchanged is shown in Fig.~\ref{vbffig}.  

The circular area of Fig.~\ref{vbffig} comprises various SM and BSM 
sub-diagrams which are added, with the result being described by 
leptonic tensors. Since these tensors are independent of the QCD 
part, in particular since they do not affect the structure of NLO 
QCD corrections, they can be modified to include arbitrary new 
physics effects and then immediately yield cross sections at NLO QCD 
accuracy: In addition to potential modifications of Higgs 
contributions, one needs to extend
the SM leptonic tensors by the additional Feynman graphs involving \spin2
particles.

\begin{figure}
\includegraphics[width=\textwidth]{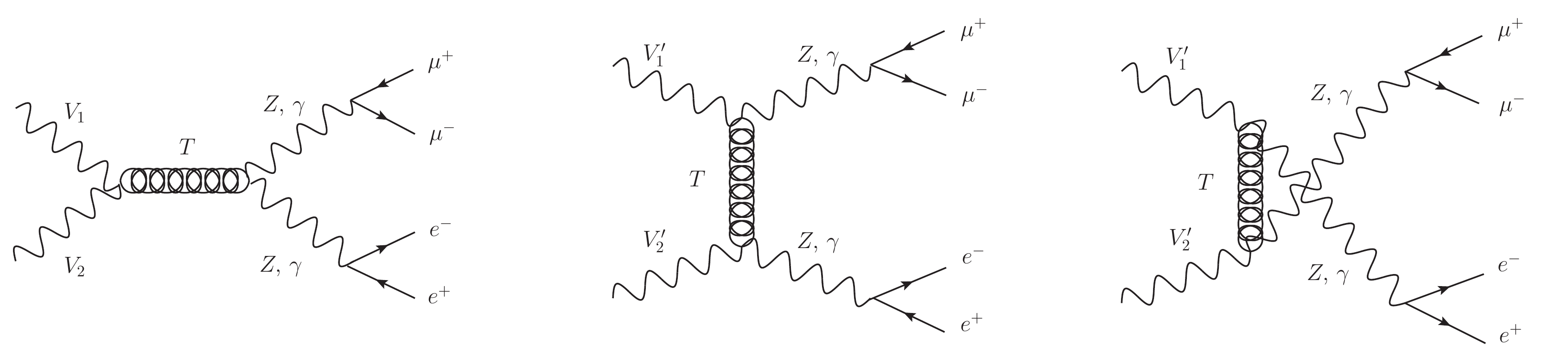}
\caption{Feynman graphs of the sub-process $V V \rightarrow e^+ \, e^-
\, \mu^+ \mu^-$ involving the \spin2 singlet particle $T$, with $V_1
V_2 \,\, \widehat{=} \,\, W^+W^-, \gamma Z, Z \gamma, \gamma \gamma, ZZ$
and $V'_1 V'_2 \,\, \widehat{=} \,\, \gamma Z, Z \gamma, \gamma \gamma,
ZZ$.}
\label{tozzll_neutr}
\end{figure}

\begin{figure}
\centerline{\includegraphics[width=0.67\textwidth]{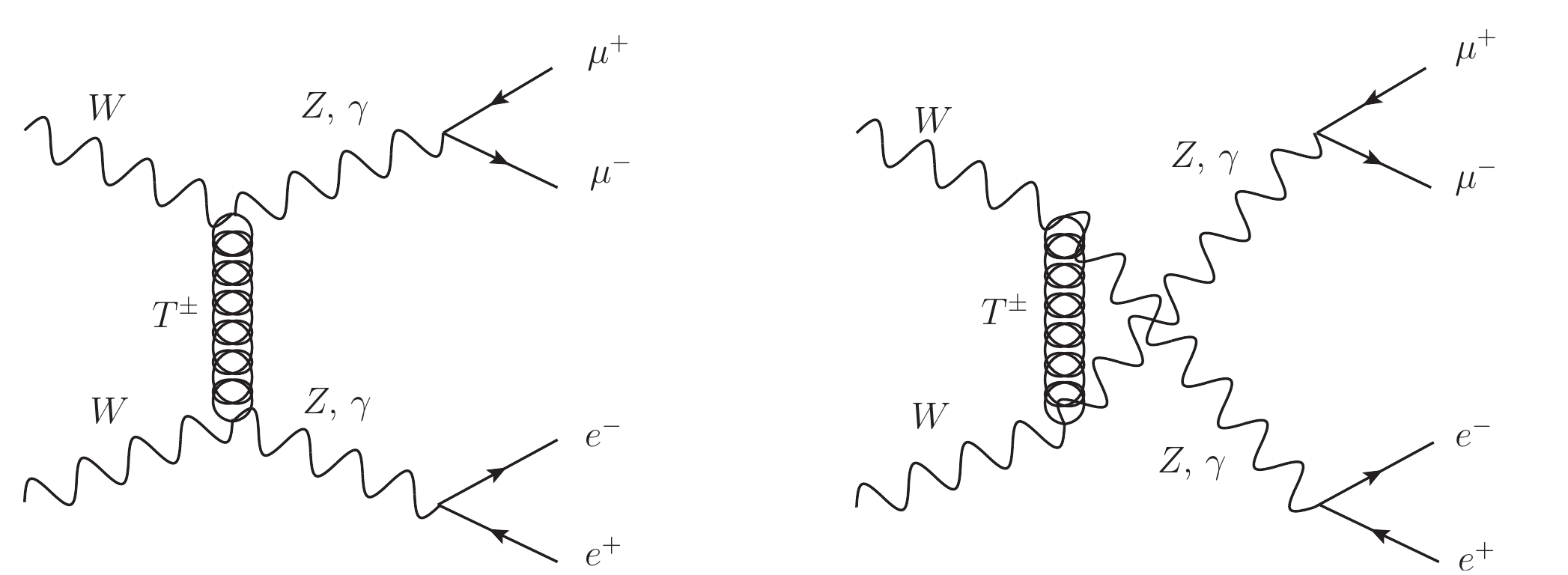}}
\caption{Feynman graphs of the sub-process $V V \rightarrow e^+ \, e^-
\, \mu^+ \mu^-$ involving charged \spin2 triplet particles.}
\label{tozzll_geladen}
\end{figure}

For $V V \rightarrow e^+ \, e^- \, \mu^+ \mu^-$, the electroweak
sub-process of $pp \rightarrow e^+ \, e^- \, \mu^+ \, \mu^- \, jj$,
these additional diagrams are depicted in Figs.\ \ref{tozzll_neutr} and
\ref{tozzll_geladen}, respectively. Fig.~\ref{tozzll_neutr} shows the
graphs involving the \spin2 singlet particle $T$.  The diagrams for the
neutral \spin2 triplet are the same as for the singlet particle, with
$T^0$ instead of $T$. The Feynman graphs for contributions of charged triplet
particles are depicted in Fig.~\ref{tozzll_geladen}.

For the other processes with four leptons and two jets in the final
state, the additional Feynman diagrams are analogous and can be found in
Ref.~\cite{eigenediplarbeit}.

Within the \textsc{Vbfnlo} program, the leptonic tensors for a given process do 
not change when going from LO to NLO contributions, nor do they change 
between quark and anti-quark initiated sub-processes. Thus they are 
calculated once per phase-space point and reused, which considerably 
speeds up the program. The Feynman diagrams contributing to the 
leptonic tensors are
calculated via calls of \textsc{Helas} routines \cite{helas}. For the
calculation of the graphs involving \spin2 particles, we created new
\textsc{Helas} routines containing the Feynman rules of the \spin2
model and also a (faster) code which directly determines the \spin2 
resonance contributions to the leptonic tensors.

\begin{figure}
\centerline{\includegraphics[width=0.67\textwidth]{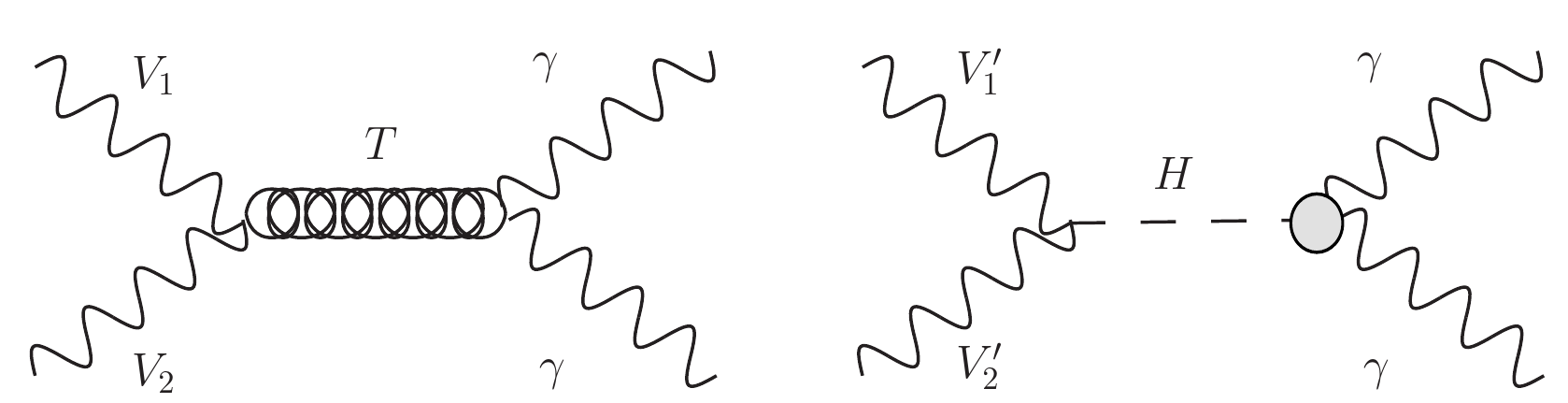}}
\caption{Feynman graphs of the sub-process $V V \rightarrow \gamma
\gamma$. Left hand side: via a \spin2 resonance (singlet or neutral triplet), with $V_1 V_2 \,\, \widehat{=} \,\, W^+W^-, \gamma Z, Z \gamma, \gamma \gamma, ZZ$. 
Right hand side: via a Higgs resonance, with $V'_1 V'_2 \,\, \widehat{=} \,\, W^+W^-, ZZ$.}
\label{figsp2toaa}
\end{figure}

In the analysis of \spin2 resonances in the VBF process 
$qq \to qq\gamma\gamma$,
we only consider sub-diagrams with resonant
production of \spin2 particles, which are shown on the left hand side of
Fig.~\ref{figsp2toaa}. Here, $T$ denotes either the singlet or the
neutral triplet particle. The SM continuum contributions are omitted, 
because interference effects with the \spin2 resonance are small 
for a narrow resonance. Thus, the continuum background 
can be eliminated via a sideband analysis, similar to the SM Higgs case
(see Sec.~\ref{sec:photonpairproduction}).  We have analyzed the non-resonant
\spin2 contributions as well, yet they were found to yield no
significant modifications. The characteristics of light \spin2
resonances in this process are compared to those of a SM Higgs
resonance.  For the analysis of the latter case, we only consider
analogous sub-diagrams with an effective $H \gamma \gamma$
coupling~\cite{Alwall:2007st}, which are depicted on the right hand side
of Fig.~\ref{figsp2toaa}. The results have been cross-checked with
another existing implementation of VBF Higgs + 2 jet production, with
the Higgs boson decaying into two photons in the narrow-width
approximation~\cite{Arnold:2011wj, Figy:2003nv}.

\begin{figure}
\includegraphics[width=\textwidth]{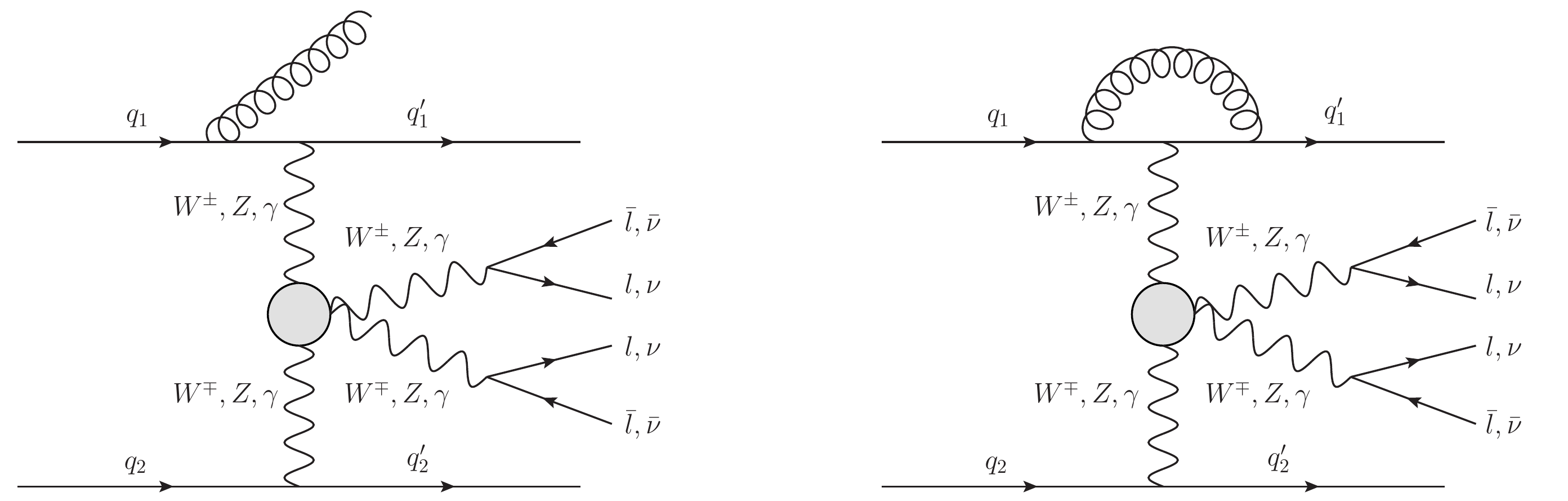}
\caption{Representative vector-boson-fusion Feynman graphs at NLO QCD. Left hand side: real emission, right hand side: virtual correction.}
\label{fignlovbf}
\end{figure}

Since the \spin2 model only affects the electroweak part of the VBF processes, the NLO QCD corrections are similar to those of the SM and can be adapted from the 
respective calculations, which are described in detail in Refs.\ \cite{Jager:2006zc} and \cite{Figy:2003nv}. The real-emission contributions are  
obtained by attaching an external gluon to the two quarks lines of Fig.~\ref{vbffig} in all possible ways, which also comprises quark-gluon initiated sub-processes. 
Due to the color-singlet structure of VBF processes, the virtual corrections only comprise Feynman diagrams with a virtual gluon attached to a single quark line. Since 
the processes with four leptons and two jets in the final state contain graphs with three electroweak bosons attached to a quark line, they contain at most pentagon 
contributions to the virtual corrections. The other graphs give rise to box, vertex and quark self-energy corrections. In the photon pair-production process, where we 
only consider graphs with a single electroweak boson attached to a quark line, the virtual corrections are much simpler, since there are no box and pentagon 
contributions. Representative diagrams for the real emission and the
virtual corrections are depicted in Fig.~\ref{fignlovbf}.

In the calculation of the NLO corrections, infrared singularities
arise both from virtual corrections and from soft and collinear
phase-space regions in the real emission part. They are canceled against
each other by using the Catani-Seymour dipole subtraction
method~\cite{Catani:1996vz}. The regularization is performed in the
dimensional-reduction scheme in $d=4-2 \epsilon$
dimensions~\cite{dimred}. For the evaluation of the finite parts of the
virtual corrections, the Denner--Dittmaier scheme~\cite{DDpentagon} is
applied for five-point functions and the Passarino--Veltman
reduction formalism~\cite{Passarino} for loop functions up to four
external legs.

Throughout the calculation, the Cabibbo--Kobayashi--Maskawa matrix
$V_{\text{CKM}}$ is approximated by the identity matrix. This gives the
same results as the exact matrix $V_{\text{CKM}}$ for the sum over all
quark flavors (as long as no final-state flavor tagging is done and mixing with the massive 
top quark is neglected), since $V_{\text{CKM}}$ is unitary.

Finite-width effects in massive vector-boson propagators are taken into
account by using a modified version~\cite{Alwall:2007st,Oleari:2003tc} of the
complex-mass scheme~\cite{Denner:1999gp}, where $m_V^2$ is replaced with
$m_V^2-i m_V \Gamma_V$, while a real value for $\text{sin}^2 \,
\theta_W$ is kept. This replacement includes the $m_V^2$ appearing in
the spin part of the propagator in the unitary gauge.
This approach is analogous to the one implemented in MadGraph~\cite{Alwall:2007st} 
and, indeed, our SM amplitudes agree with the ones obtained with MadGraph. \\
In the full complex-mass scheme the SM amplitudes are gauge invariant. The BSM 
contributions appear as $s$-, $t$- and/or $u$- channel \spin2 exchange graphs 
in weak-boson scattering with a single \spin2 propagator. Since they are derived 
from the gauge invariant Lagrangians (\ref{spin2lagrangian}) or (\ref{tripletlagrangian}), 
the resulting amplitudes are gauge invariant in the absence of finite-width effects. 
One might worry that using the finite-width propagator (\ref{propagator1}) for the 
\spin2 fields might break electroweak gauge invariance. We have checked, however, that 
changing to the overall-factor scheme (which respects gauge invariance), i.e. removing
the width from all \spin2 propagators and multiplying the total BSM
amplitude with a factor $\frac{(p^2-m^2)}{(p^2-m^2+i m \Gamma)}$, leaves our
results unchanged within the numerical accuracy. Here \mbox{$p$, $m$ and $\Gamma$}
denote the momentum, mass and width of the s-channel \spin2 particle,
respectively. While our approach to the handling of finite-width effects 
is not fully gauge invariant, a comparison with fully gauge invariant amplitudes 
shows that this does not lead to noticeable deviations at LHC energies. 
\footnote{For processes which include graphs with two or more \spin2 particles, 
the propagator of Eq.~(\ref{propagator1}) would require further modification 
to insure gauge invariance. This complication does not arise in the present context.}

\subsection{\label{sec:settings}Input parameters and selection cuts}

As electroweak input parameters, we choose $m_W = 80.399$ GeV, $m_Z =
91.1876$ GeV and $G_F = 1.16637 \cdot 10^{-5} \,\text{GeV}^{-2}$, which
are taken from the 2010 results of the Particle Data Group~\cite{PDG}. 
$\alpha$ and $\sin^2 \theta_W$ are derived from these quantities using
tree-level electroweak relations. 
We use the CTEQ6L1~\cite{Pumplin:2002vw} parton distribution functions
at LO and the CT10~\cite{Lai:2010vv} set at NLO with
$\alpha_s(m_Z)=0.118$.
The factorization scale and the renormalization scale are set to $\mu_F
= \mu_R = Q = \sqrt{|q_{if}^2|}$, where $q_{if}$ is the 4-momentum
transfer between the respective initial and final state quarks. With 
this scale choice, LO calculations were found to give a good
approximation of NLO cross sections and distributions, while
the NLO results are hardly sensitive to the scale choice
\cite{Bozzi:2007ur}.  Jets are recombined from the final state partons
by using the $k_{\bot}$ jet finding algorithm~\cite{Seymour:1997kj}.

Vector-boson-fusion events are characterized by two tagging
jets in the forward regions, with decay products of the vector bosons
lying in the central-rapidity region between them. 
By applying the following inclusive VBF cuts, these features can be used to
improve the signal-to-background ratio.
The two tagging jets (i.e.\ the two jets of highest transverse momentum)
are supposed to lie inside the rapidity range accessible to the detector
and to have large transverse momenta: 
\begin{equation}
 p_{T,j}^{\text{tag}} > 30 \, \text{GeV}, \hspace{1cm} |\eta_j| < 4.5.
\end{equation}
They are reconstructed from massless partons of pseudorapidity $|\eta| <
5$ and have to be well separated: 
\begin{equation}
\Delta R_{jj} \equiv \sqrt{(\eta_{j1}-\eta_{j2})^2 + (\phi_{j1}-\phi_{j2})^2} > 0.7.
\end{equation}
In order to reduce backgrounds and make use of the characteristic
features of the VBF channel, we require a large rapidity separation of the tagging jets  
\begin{equation}
\Delta \eta_{jj} > 4,
\end{equation}
which have to be located in opposite detector hemispheres, 
\begin{equation}
\eta_{j1}^{\text{tag}} \times \eta_{j2}^{\text{tag}} < 0
\end{equation}
and have a large invariant mass
\begin{equation}
m_{jj} > m_{jj}^{\text{min}},
\end{equation}
where $m_{jj}^{\text{min}} = 1000$ GeV for the process $pp \rightarrow e^+ \, \nu_e \, \mu^- \, \overline{\nu}_\mu \, jj$ (i.e.\ ``$W^+W^-jj$ production'')
and $m_{jj}^{\text{min}} = 500$ GeV for all other processes considered in the present analysis.

The charged decay leptons (or decay photons, respectively) are required
to be hard, located at central rapidities and well-separated from the
jets: 
\begin{equation}
p_{T,l} > 20 \, \text{GeV}, \hspace{1cm} |\eta_l| < 2.5, \hspace{1cm} \Delta R_{lj} > 0.4. 
\end{equation}
Furthermore, they are supposed to fall into the rapidity gap between the two tagging jets: 
\begin{equation}
\eta_{j,\text{min}}^{\text{tag}} < \eta_l < \eta_{j,\text{max}}^{\text{tag}}.
\end{equation}
Here, $l$ denotes a charged lepton or a photon, depending on the considered process.
In order to have isolated photons, we require an additional minimal photon-photon \mbox{$R$-separation} 
\begin{equation}
\Delta R_{\gamma \gamma} > 0.4 
\end{equation}
and impose photon isolation from hadronic activity as suggested in 
Ref.~\cite{Frixione:1998jh}
with separation parameter $\delta_0=0.7$, efficiency
$\epsilon=1$ and exponent $n=1$.

To avoid singularities from quasi-collinear $\gamma^* \rightarrow l^+ l^-$ decays, we apply a cut on the invariant mass of two same-flavor charged leptons, 
\begin{equation}
m_{ll} > 15 \, \text{GeV}.
\end{equation}

By imposing this set of cuts, the LO differential cross sections are
finite, since they lead to finite scattering angles for the two jets. In
the NLO calculation, initial state singularities arise, resulting from
collinear quark and gluon splittings ($q \rightarrow q g$ and $g
\rightarrow q \bar{q}$). 
They are factorized into the PDFs. 
Moreover, divergences from
$t$-channel exchange of photons with low virtuality appear 
in the real-emission part, 
when the additional parton radiation is resolved as a separate jet, but
for the other, non-radiating, quark line the initial- and final-state quarks
become collinear. These divergences are of electroweak origin and could 
be removed by including a photon density in the proton PDFs. 
However, the precise treatment of these
divergences does not appreciably affect the cross section, in particular
when VBF cuts are applied~\cite{Oleari:2003tc}. Therefore, we eliminate them by imposing an 
additional cut on the photon virtuality,
\begin{equation}
Q^2_\gamma > 4 \, \text{GeV}^2.
\end{equation}

When distributions of a \spin2 resonance in the processes with four
leptons and two jets in the final state are studied, it is convenient to
cut off contributions which do not stem from the resonance.
 Therefore, a minimal and a maximal invariant mass of the final-state
lepton system can be required. Whenever such a cut is applied, it will
be specified in the corresponding section.

\FloatBarrier
\section{Results \label{sec:Results}}

\FloatBarrier
\subsection{VBF photon pair-production via Higgs or Spin-2 resonances \label{sec:photonpairproduction}}

In this section, we present numerical results of our analysis of Higgs
and \spin2 resonances in VBF photon pair-production in association with
two jets.  We list the cross sections at LO and NLO QCD accuracy,
discuss transverse-momentum distributions and the relevance of the
formfactor (\ref{formfactor}) and present angular distributions. 
Thereby, we investigate whether a \spin2 and a SM Higgs resonance can be
distinguished from one another. Additionally, different parameter
settings of the \spin2 model are studied and the \spin2 singlet is
compared to the \spin2 triplet case. Furthermore, the impact of the NLO 
QCD corrections is analyzed.

If not indicated otherwise, we consider a \spin2 singlet resonance with
couplings \mbox{$f_1=0.04, f_2=0.08, f_5 = 10,$} $f_{i\neq 1,2,5} = 0$ and 
$\Lambda = 21\, \text{TeV}$.
The parameters of the formfactor are $\Lambda_{ff} = 400 \, \text{GeV},\,
n_{ff} = 3$. These parameters are chosen in order to approximately reproduce
 the relative branching ratios for a SM Higgs boson of mass 126~GeV decaying 
 into $\gamma\gamma$, $WW$ and $ZZ$ as
 well as the SM predictions for the transverse momentum distributions of the
 tagging jets in VBF Higgs boson production at the LHC (see below).
For the triplet case, we use the same parameters for the formfactor settings,
but set the couplings to $f_6 = f_7 = 1, f_{i\neq 6,7} = 0,\, \Lambda = 9\,
\text{TeV}$. 
The mass of the Higgs boson and the \spin2 particles is set to 126 GeV 
and we assume $pp$ collisions at a centre of mass energy of 8 TeV. 
In the following, when figures compare different values of coupling parameters,
couplings $f_i$ which are not given explicitly are set to zero and
$\Lambda$ is adjusted such that the cross section is approximately the 
same as the one of the SM Higgs resonance. 
If not indicated otherwise, differential distributions are determined 
in the laboratory frame.

In order to analyze possible effects of a finite detector resolution, a
smearing of the energy and the transverse momenta of the final-state
photons and jets according to a Gaussian distribution was performed,
which is based on a CMS Monte-Carlo study~\cite{CMSnote}. 
However, this smearing was found to have no significant influence on the
distributions we studied. Therefore, the results which are presented
here were performed without smearing. 

Table~\ref{crosssectionsphoton} gives a comparison of the integrated cross
sections for Higgs and \spin2 resonances in the photon pair-production
process at LO and NLO QCD accuracy. 
Numbers are shown for the LHC at a centre of mass energy of 8 TeV as
well as for 14 TeV. The shape of the distributions, which we will show
in the next part, is identical in both cases, so we will restrict
ourselves to the 8 TeV case there.
The $K$-factor is defined as
$K=\sigma_{\text{NLO}} / \sigma_{\text{LO}}$.
The statistical errors from the Monte Carlo integration are around one per mill 
for all the different parameter settings we study in this
section. Due to the scale choice $\mu_F = \mu_R = Q$ (see 
Sec. \ref{sec:settings}), the NLO corrections are quite small. They are roughly
the same for a Higgs and a \spin2 resonance.

\begin{table}
\begin{center}
\begin{tabular}{|l|c|c|c|c|c|c|}
\hline
 & \multicolumn{2}{|c|}{LO cross section [fb]} 
 & \multicolumn{2}{|c|}{NLO cross section [fb]} 
 & \multicolumn{2}{|c|}{$K=\frac{\sigma_{\text{NLO}}}{\sigma_{\text{LO}}}$} \\
\hline
$\sqrt{S}$ & 8 TeV & 14 TeV & 8 TeV & 14 TeV & 8 TeV & 14 TeV \\
\hline\hline
Higgs          & 0.7348(3) & 2.179(1) & 0.7448(4) & 2.241(1) & 1.014 & 1.028 \\
\hline
Spin-2 singlet & 0.7711(4) & 2.409(1) & 0.7878(4) & 2.495(1) & 1.022 & 1.036 \\
\hline
Spin-2 triplet & 0.7927(4) & 1.969(1) & 0.8041(5) & 2.098(1) & 1.014 & 1.066 \\
\hline
\end{tabular}
\caption{Integrated cross sections for different types of resonances at LO and NLO QCD accuracy for VBF 
photon pair-production. The cuts of Section \ref{sec:settings} are applied.} \label{crosssectionsphoton}
\end{center}
\end{table}

The width of the Higgs resonance is only $\approx 4$ MeV, whereas the width of the \spin2 resonance is even much smaller
 for all the different parameter settings studied here and does not exceed $0.05$ MeV (see App.~\ref{sec:decaywidths}
for details).
In principle, the width of the \spin2 resonance can be adjusted to the
one of the Higgs by choosing a smaller value of the branching ratio
parameter $b$, which quantifies the amount of additional, possibly
hard to detect, decay modes of the \spin2 particle, such as $T\to gg$ which 
could be induced by a effective $Tgg$ coupling analogous to the 
$f_1$ and $f_2$ terms in Eq.~(\ref{spin2lagrangian}).
At the same time, $\Lambda$ can be tuned such that the cross section
remains comparable to the Higgs case. 
However, the
width of the resonance peak in the di-photon mass spectrum is dominated
by the experimental resolution, which is about a GeV for CMS and ATLAS.
Therefore, these details do not play any role.

\subsubsection{Transverse-momentum distributions and formfactor \label{sec:ptdistributions}}
\FloatBarrier

\begin{figure}
\vspace{1.5cm}
 \begin{minipage}{0.5\textwidth}%
		\includegraphics[trim=30mm 20mm 70mm 80mm, width=\textwidth]{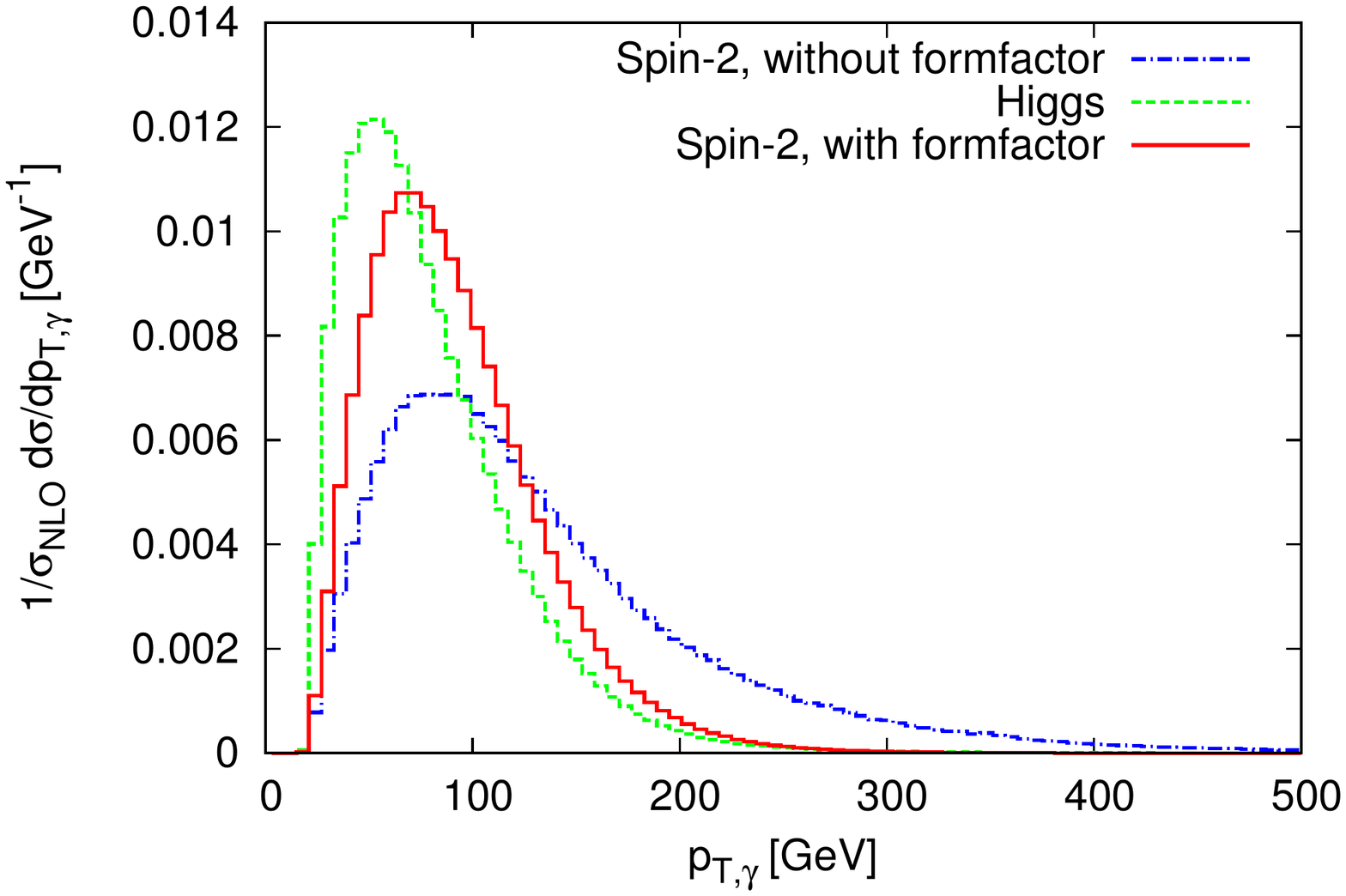}
	\end{minipage}
	\begin{minipage}{0.5\textwidth}%
		\includegraphics[trim=20mm 20mm 80mm 80mm, width=\textwidth]{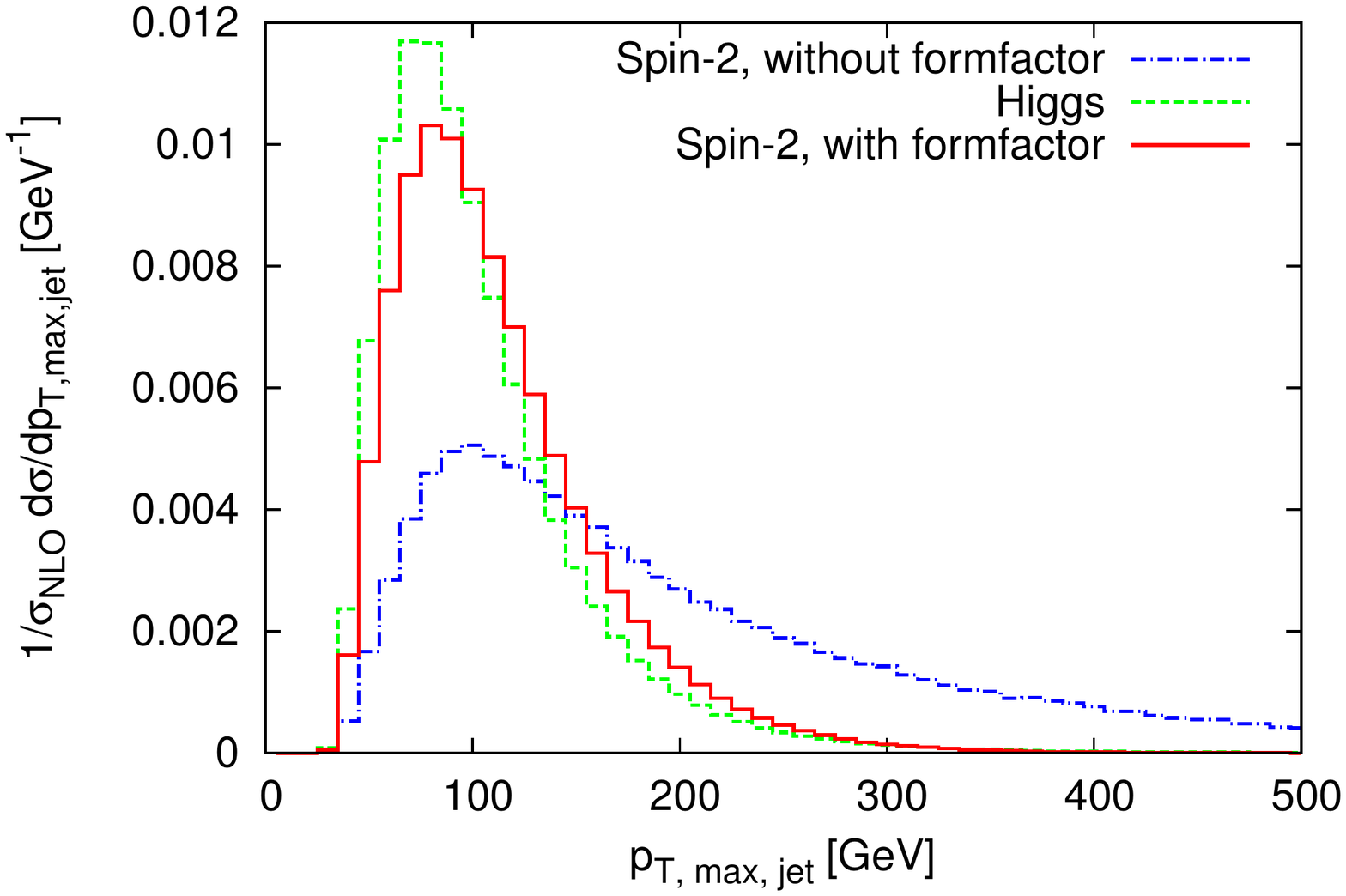}
	\end{minipage} 
\vspace{-0.2cm}
\caption{Transverse-momentum distributions for a Higgs and for a \spin2 resonance with couplings $f_1=0.04, f_2=0.08, f_5 = 10, f_{i\neq 1,2,5} = 0$, 
with and without formfactor, 
 at NLO QCD accuracy. Left hand side: $p_T$ of a final-state photon, right hand side: $p_T$ of the tagging jet with the largest transverse momentum.}
\label{pt_formfac}
\end{figure}

Fig.~\ref{pt_formfac} depicts the normalized transverse-momentum
distributions of a final-state photon and of the tagging jet with the
largest transverse momentum for a Higgs and a \spin2 singlet resonance
with and without the formfactor (\ref{formfactor}) at NLO QCD accuracy.
For a \spin2 resonance without the formfactor (or with $n_{ff}=0$ or
$\Lambda_{ff} \rightarrow \infty$, respectively), the transverse momenta
of the photons and the hardest jet are much higher than for a Higgs, so
that both cases could be easily distinguished from one another via the $p_T$
distributions.  However, the harder $p_T$ distributions for the \spin2 case 
are mainly due to the higher energy dimensions of the couplings in the 
effective Lagrangians (\ref{spin2lagrangian}) and (\ref{tripletlagrangian}) 
rather than being a measure of the spin of the resonance.  
Furthermore, unitarity of the
S-matrix in elastic weak-boson scattering is violated for the present
\spin2 model if no formfactor  is applied (for more details, see Ref.\
\cite{eigenediplarbeit}). By a judicious choice of the formfactors,
e.g. Eq.~(\ref{formfactor}) with 
$\Lambda_{ff} = 400 \, \text{GeV},\, n_{ff} =3$, the $p_T$
distributions of the \spin2 resonance can be adjusted to closely resemble 
those of the SM Higgs boson. 
This works for the transverse momenta of the photons and the jets
simultaneously (see Fig.\ \ref{pt_formfac}). Therefore, 
transverse-momentum distributions which look like those of
the Higgs would not be a proof for a Higgs resonance. These
distributions could originate from a \spin2 resonance with an adequate
formfactor as well. An analogous formfactor study has been performed
previously concerning anomalous Higgs couplings~\cite{hep-ph/0403297}.
From now on, the formfactor parameters are set to $\Lambda_{ff} = 400 \,
\text{GeV},\, n_{ff} = 3$ throughout this subsection. 

\begin{figure}
\vspace{1.5cm}
 \begin{minipage}{0.5\textwidth}%
		\includegraphics[trim=30mm 20mm 70mm 80mm, width=\textwidth]{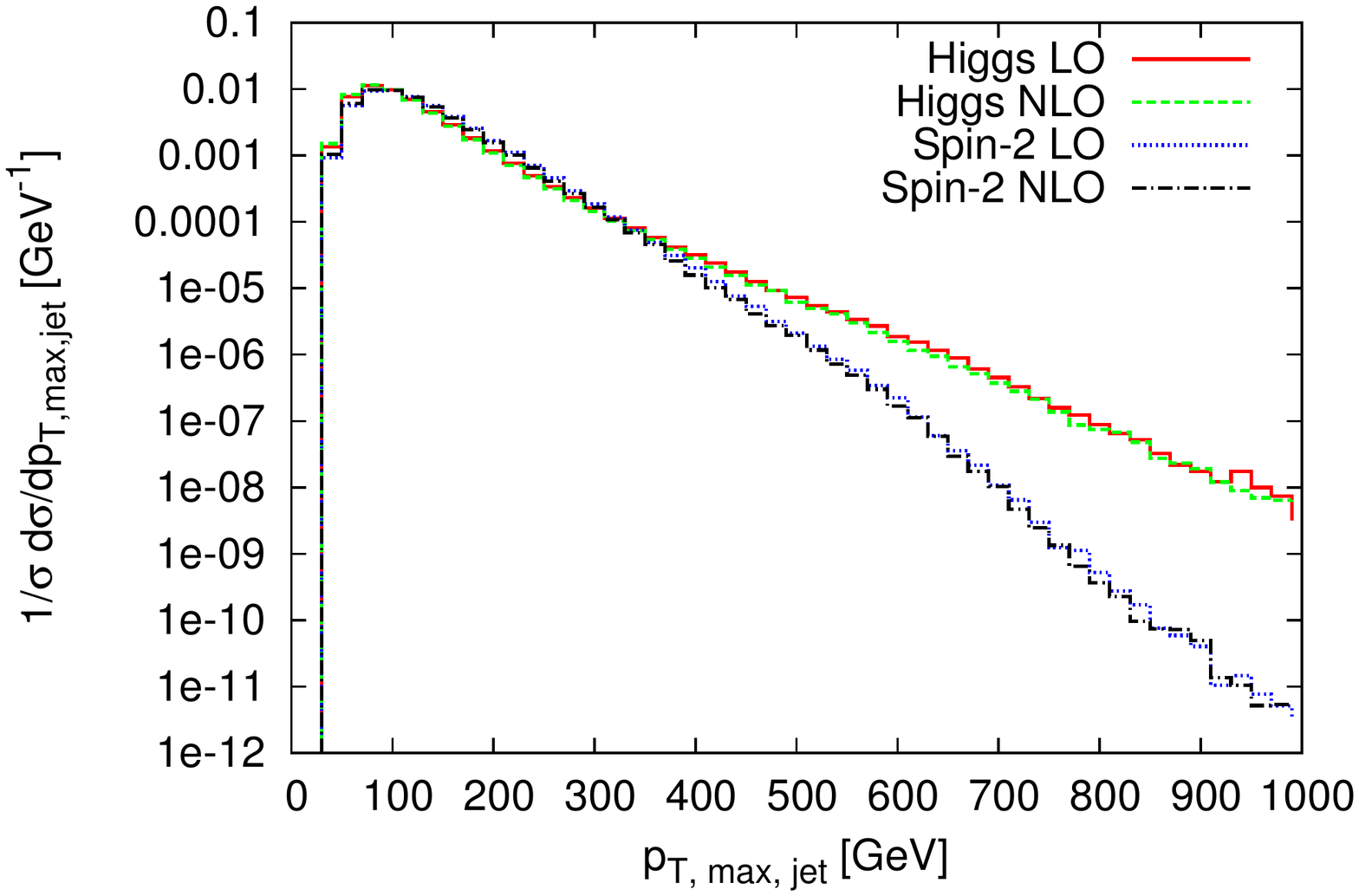}
	\end{minipage}
	\begin{minipage}{0.5\textwidth}%
		\includegraphics[trim=20mm 20mm 80mm 80mm, width=\textwidth]{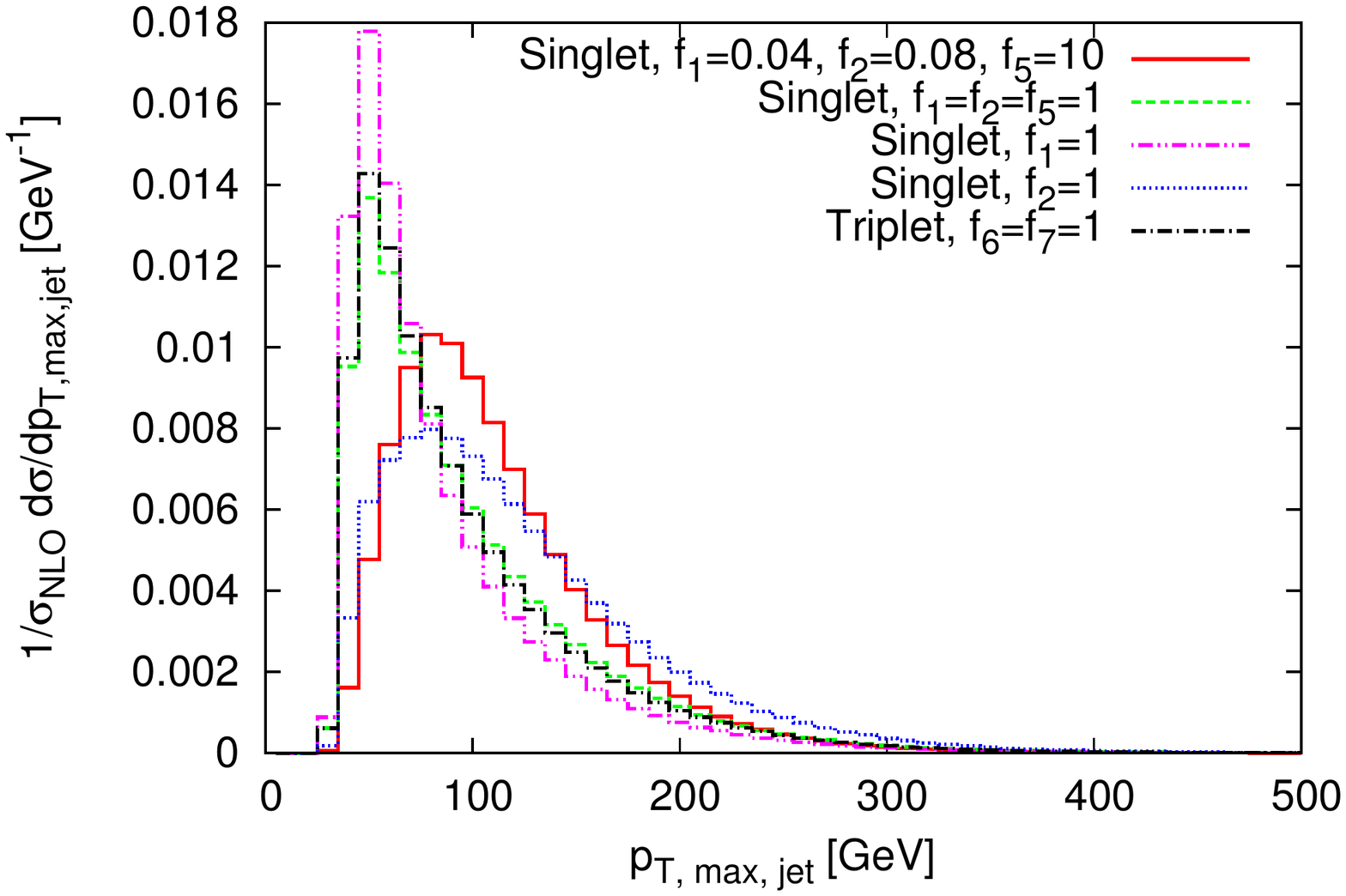}
	\end{minipage} 
\vspace{-0.2cm}
\caption{Normalized $p_T$ distribution of the tagging jet with the largest
  transverse momentum. Left hand side: Higgs and \spin2 resonance (with
  $f_1=0.04, f_2=0.08, f_5 = 10, f_{i\neq 1,2,5} = 0$) at LO and NLO QCD
  accuracy. Right hand side: Spin-2 singlet and triplet resonance with
  different coupling parameters at NLO QCD accuracy.} \label{pt_verschpara}
\end{figure}

On the left hand side of Fig.~\ref{pt_verschpara}, the impact of the NLO
corrections on the transverse momentum of the hardest jet is
illustrated. In order to compare the shape, LO distributions are
normalized to the LO cross section and NLO distributions to the NLO
cross section. 
The NLO corrections
tend to shift the distributions to smaller values of $p_T$, since a
fraction of the total transverse momentum is carried by the additional
gluon in the real emission contribution. This feature is analogous to
the SM case \cite{Jager:2006zc, Figy:2003nv} and independent of the spin
of the resonance. For the \spin0 and \spin2 case, this is depicted in
Fig.~\ref{pt_verschpara}, while an analogous plot
for \spin1 can be found in Ref.~\cite{Englert:2008}. Due to our scale choice, 
the impact of the NLO corrections is small, 
as it is for the integrated cross section as well (see
Table~\ref{crosssectionsphoton}). While the $K$-factor in the high $p_T$
region  ($400 \, \text{GeV} < p_{\text{T, max, jet}} < 900 \, \text{GeV}$) is
around 0.9 for the spin-2 case with $\mu_F = \mu_R = Q$, it would be around
0.6 if we had chosen $\mu_F = \mu_R = m_W$ instead, mainly because of a higher
prediction for the LO cross section.

The $p_T$ distributions of a \spin2 resonance depend slightly on the
coupling parameters, which is exemplified on the right hand side of
Fig.~\ref{pt_verschpara} for the transverse momentum of the hardest jet. 
This can be understood from the Feynman rules (Eq. \ref{Feynmanrules}): For \mbox{$f_1=1, f_{i\neq 1} = 0$}, 
spin-2 resonances are mainly produced by initial photons, which leads to an enhancement of the low $p_T$ region, 
while for the cases $f_2=1, f_{i\neq 2} = 0$ and $f_1=0.04, f_2=0.08, f_5 = 10$, initial $W$ and $Z$ bosons dominate.  
It can also be seen that the $p_T$ distributions of the \spin2 triplet
resonance can be adjusted to those of the Higgs with appropriate formfactor
settings.

\FloatBarrier
\subsubsection{Angular distributions}

In this section, various angular distributions are compared for a Higgs
and a \spin2 resonance and for different \spin2 couplings. Furthermore,
the impact of the NLO QCD corrections is illustrated. If not indicated
otherwise, distributions are presented at NLO QCD accuracy.  Note that
all the figures include a normalization factor
$1/\sigma_{\text{NLO}}$.

\begin{figure}[H]
\vspace{1cm}
\centerline{\includegraphics[trim=25mm 20mm 75mm 70mm, width=0.5\textwidth]{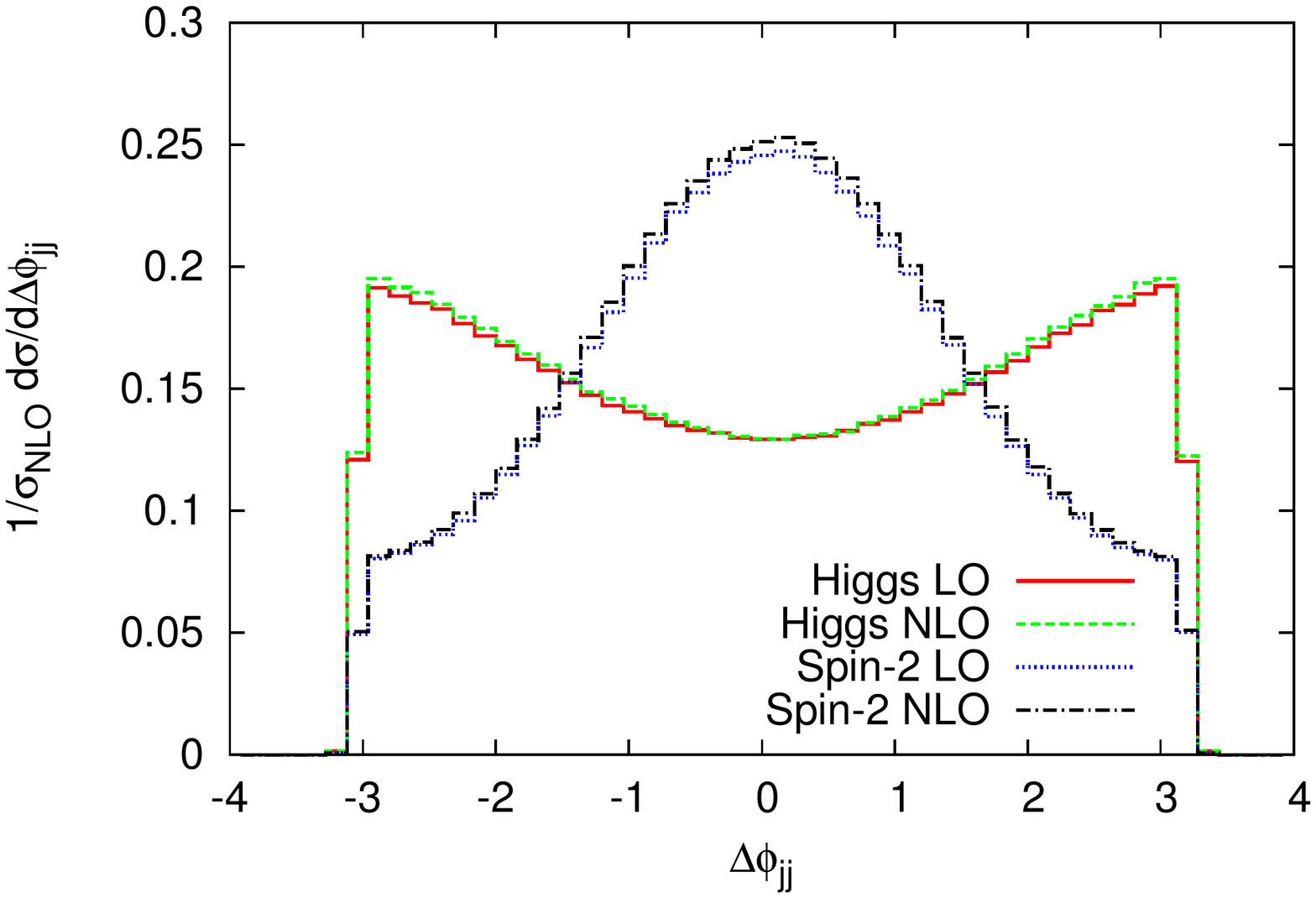}}
\caption{Azimuthal angle difference between the two tagging jets for a
Higgs and for a \spin2 resonance with couplings $f_1=0.04, f_2=0.08, f_5 = 10, f_{i\neq 1,2,5} = 0$, 
both at LO and NLO QCD accuracy.} \label{deltaphijj}
\end{figure}

\begin{figure}[H]
 \begin{minipage}{0.5\textwidth}%
		\includegraphics[trim=30mm 20mm 70mm 50mm, width=\textwidth]{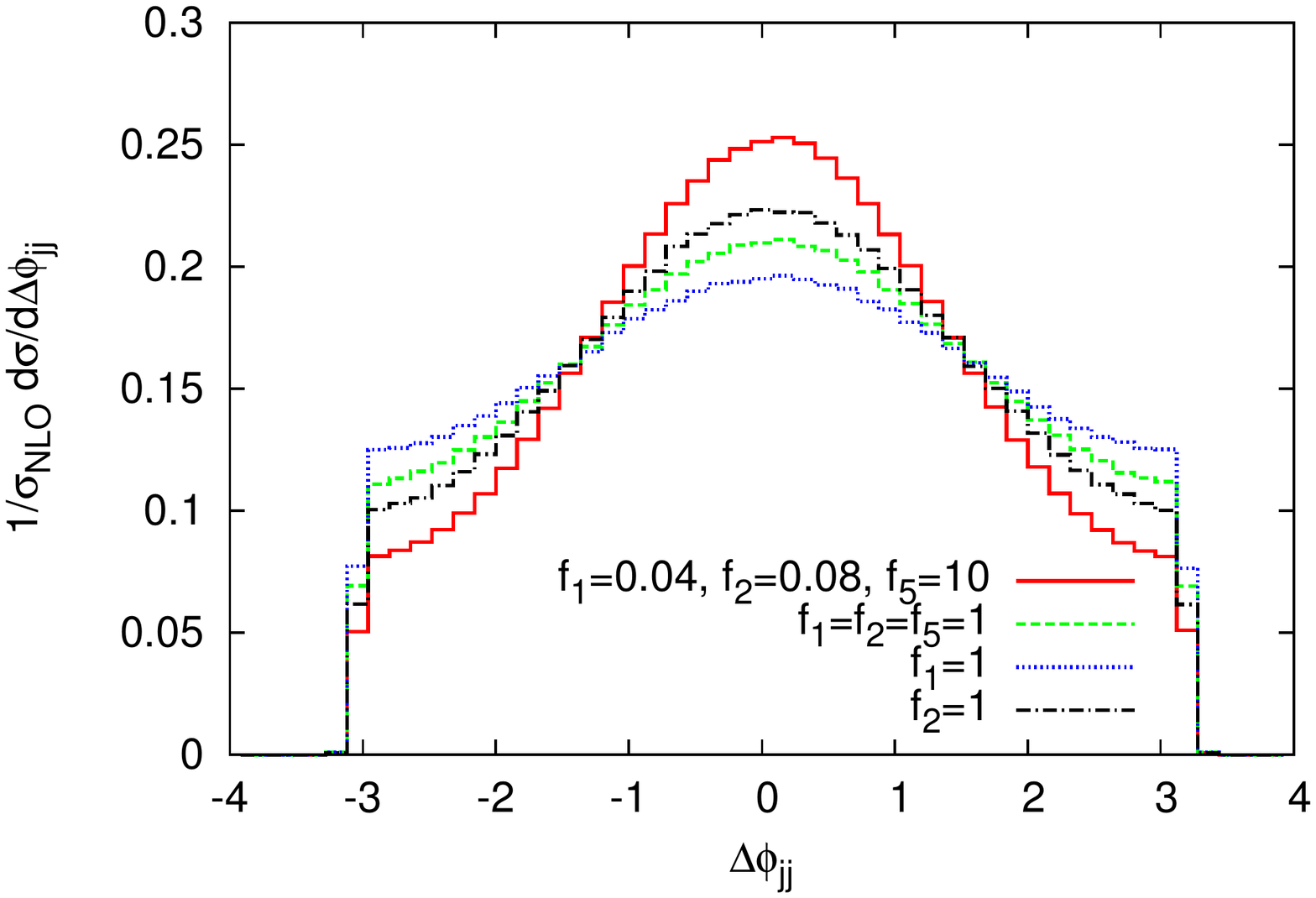}
	\end{minipage}
	\begin{minipage}{0.5\textwidth}%
		\includegraphics[trim=20mm 20mm 80mm 50mm, width=\textwidth]{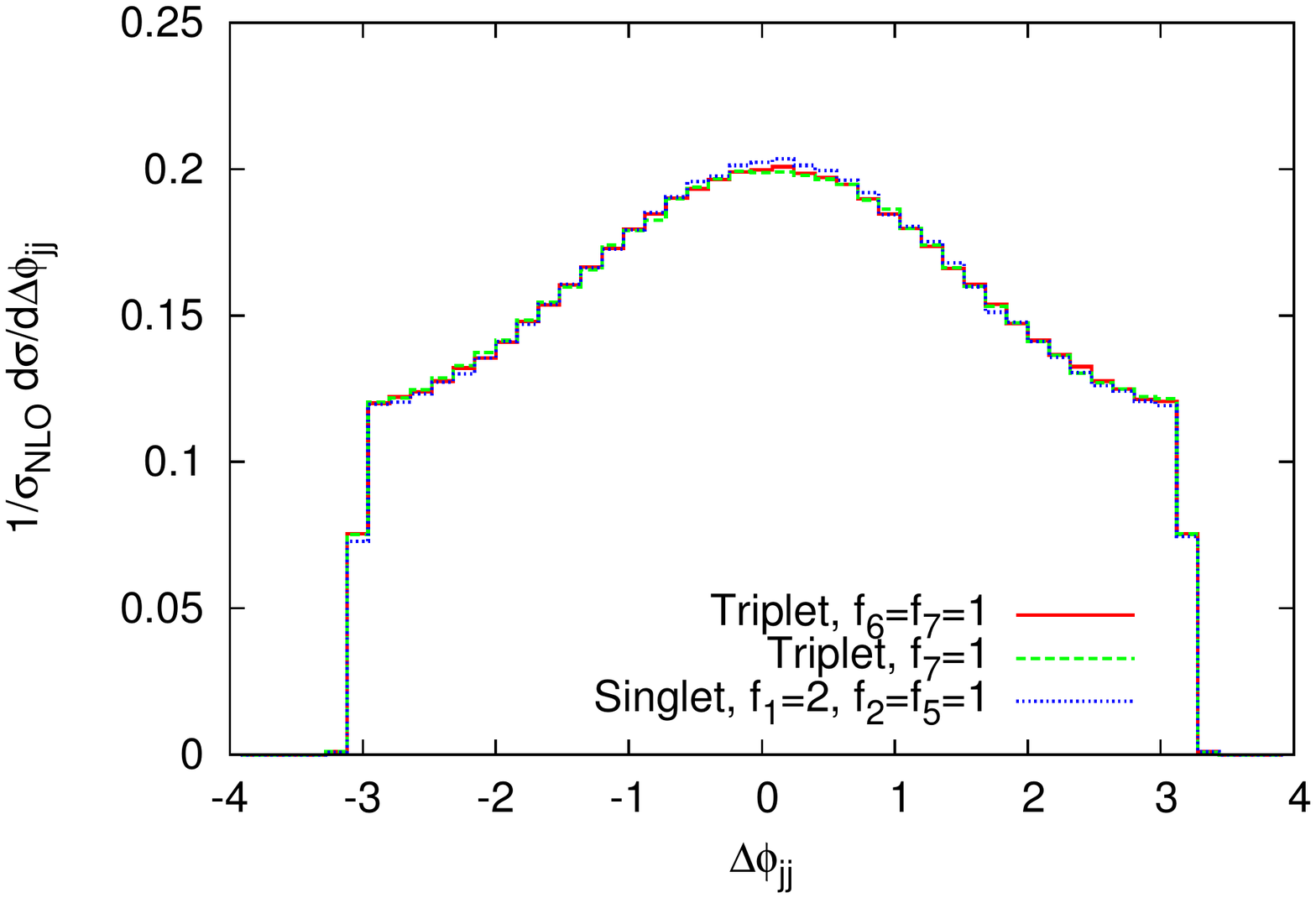}
	\end{minipage} 
\caption{Azimuthal angle difference between the two tagging jets for a
\spin2 resonance with different coupling parameters at NLO QCD accuracy. Left hand side: 
\spin2 singlet, right hand side: \spin2 singlet and triplet.} \label{deltaphijj_verschpara}
\end{figure}

First, we analyze the azimuthal angle difference between
the two tagging jets. This observable has the capability to distinguish between different 
structures of $HVV$ couplings ~\cite{HiggsCP, hep-ph/0403297}. 
Fig.\ \ref{deltaphijj} depicts the respective distribution for a SM Higgs and a \spin2 singlet resonance at LO
and NLO QCD accuracy. In both cases, the characteristic shape of the
distribution is not modified by the NLO corrections, the curves are just
slightly shifted according to the overall $K$-factor (see Table
\ref{crosssectionsphoton}).\\
Different \spin2 couplings lead to a slightly different $\Delta
\Phi_{jj}$ distribution of a \spin2 singlet resonance (left hand side of 
Fig.\ \ref{deltaphijj_verschpara}), 
yet its characteristic shape is nearly independent of these parameters. 
As shown on the right hand side of Fig.\ \ref{deltaphijj_verschpara},
the $\Delta \Phi_{jj}$ distribution of the \spin2 triplet case resembles the 
singlet case with coupling parameters 
$f_1=2, f_2=f_5=1, f_{i\neq1,2,5}=0$.
Different settings of the triplet couplings hardly influence the
distribution, since the contribution of $f_6$ is negligible.
In fact, this is the case for all angular distributions considered. 
All in all, the $\Delta \Phi_{jj}$ distribution shows a fundamental
difference between a Higgs and a \spin2 resonance, which is nearly
independent of the \spin2 coupling parameters. Therefore, it provides an
important tool to distinguish between the two resonances.

\begin{figure}[H]
\vspace{1.5cm}
\centerline{\includegraphics[trim=25mm 20mm 75mm 80mm, width=0.5\textwidth]{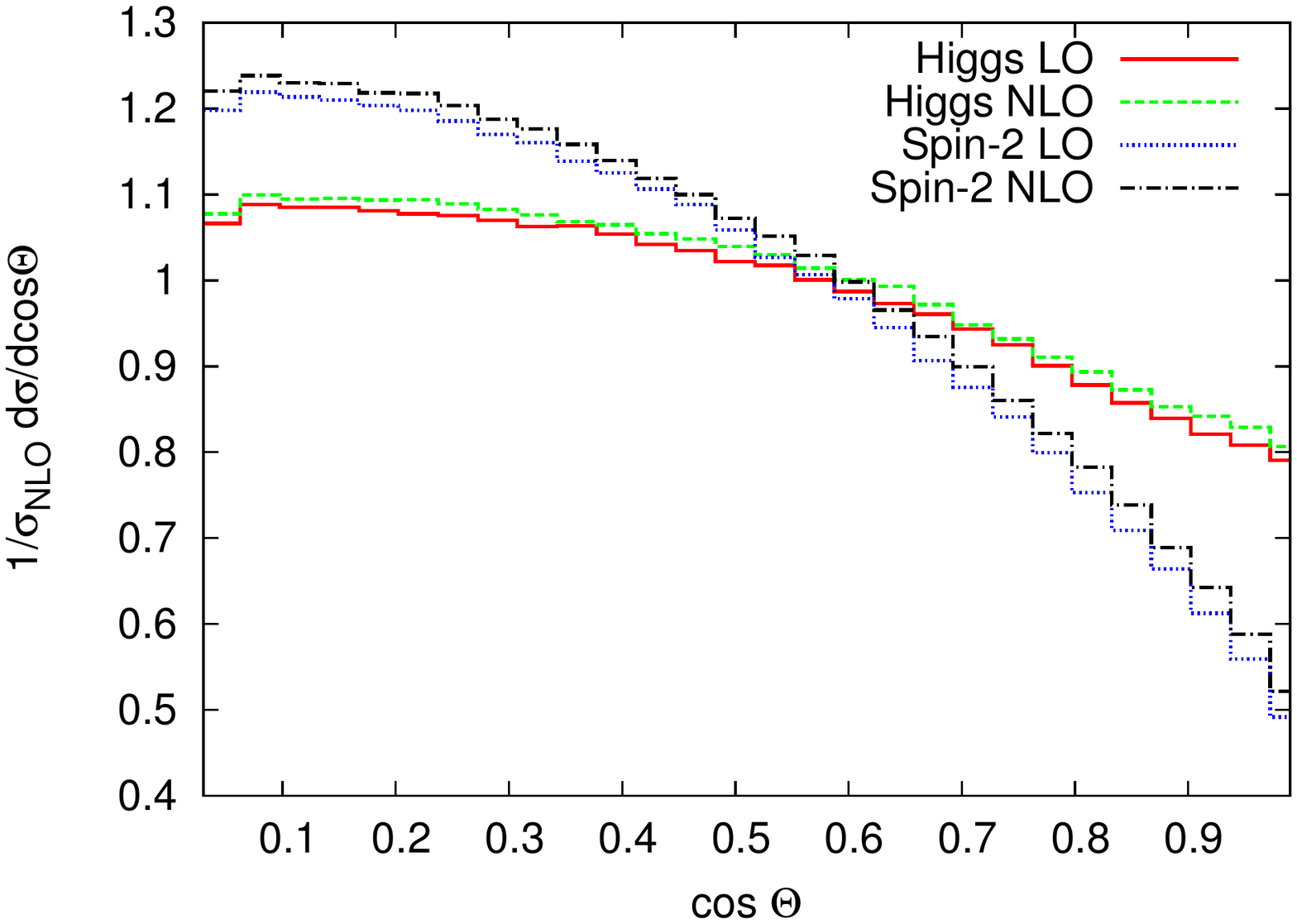}}
\caption{Normalized $\cos \Theta$ distribution for a Higgs and for a \spin2 resonance with couplings $f_1=0.04, f_2=0.08,$ \mbox{$f_5 = 10,$} $f_{i\neq 1,2,5} = 0$, 
both at LO and NLO QCD accuracy. $\Theta$ is the angle between the reconstructed momentum of an initial electroweak boson and an 
outgoing photon in the rest frame of the resonance.}\label{costhetav1}
\end{figure}

\begin{figure}[H]
\vspace{1.5cm}
 \begin{minipage}{0.5\textwidth}%
		\includegraphics[trim=30mm 20mm 70mm 80mm, width=\textwidth]{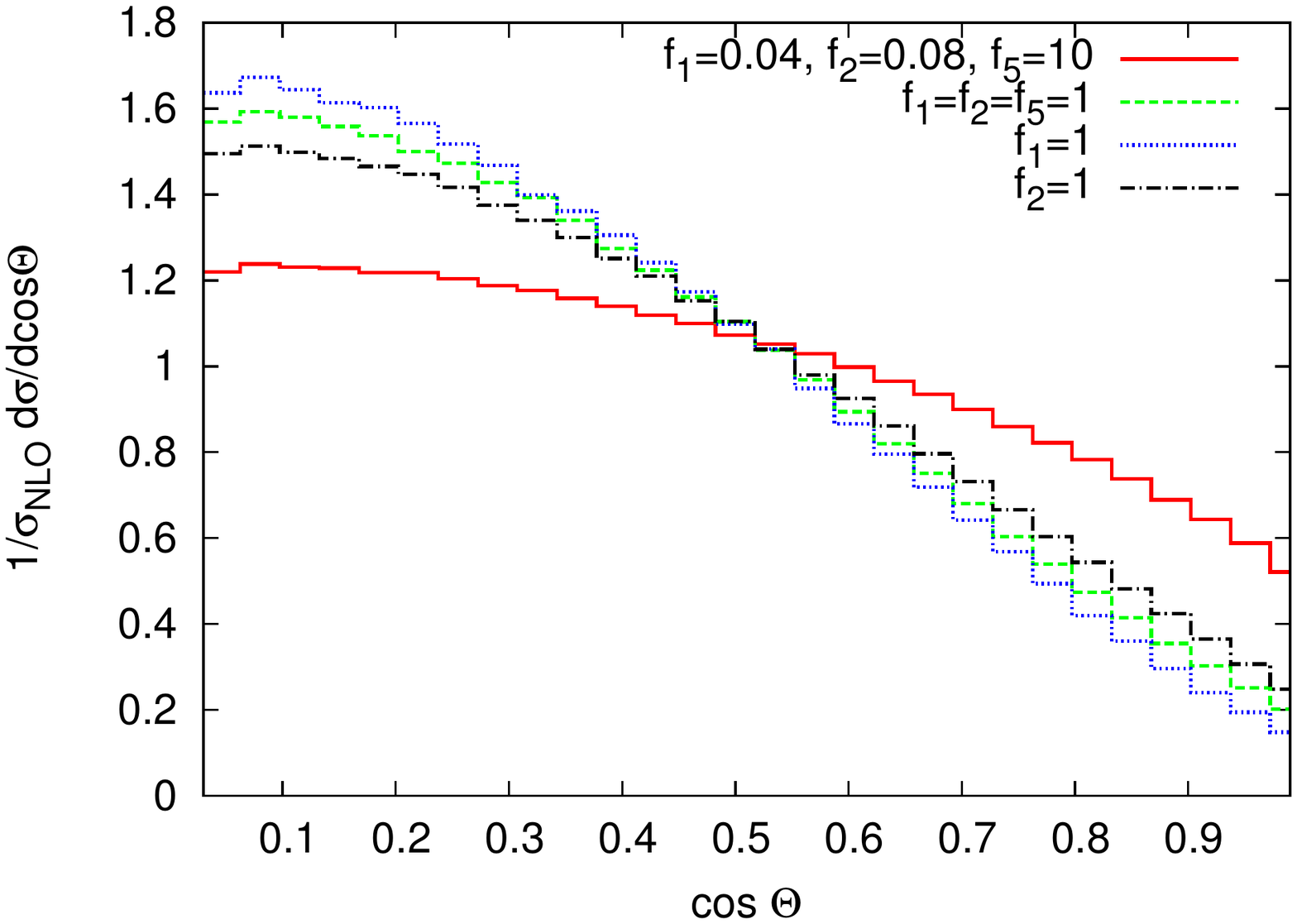}
	\end{minipage}
	\begin{minipage}{0.5\textwidth}%
		\includegraphics[trim=20mm 20mm 80mm 80mm, width=\textwidth]{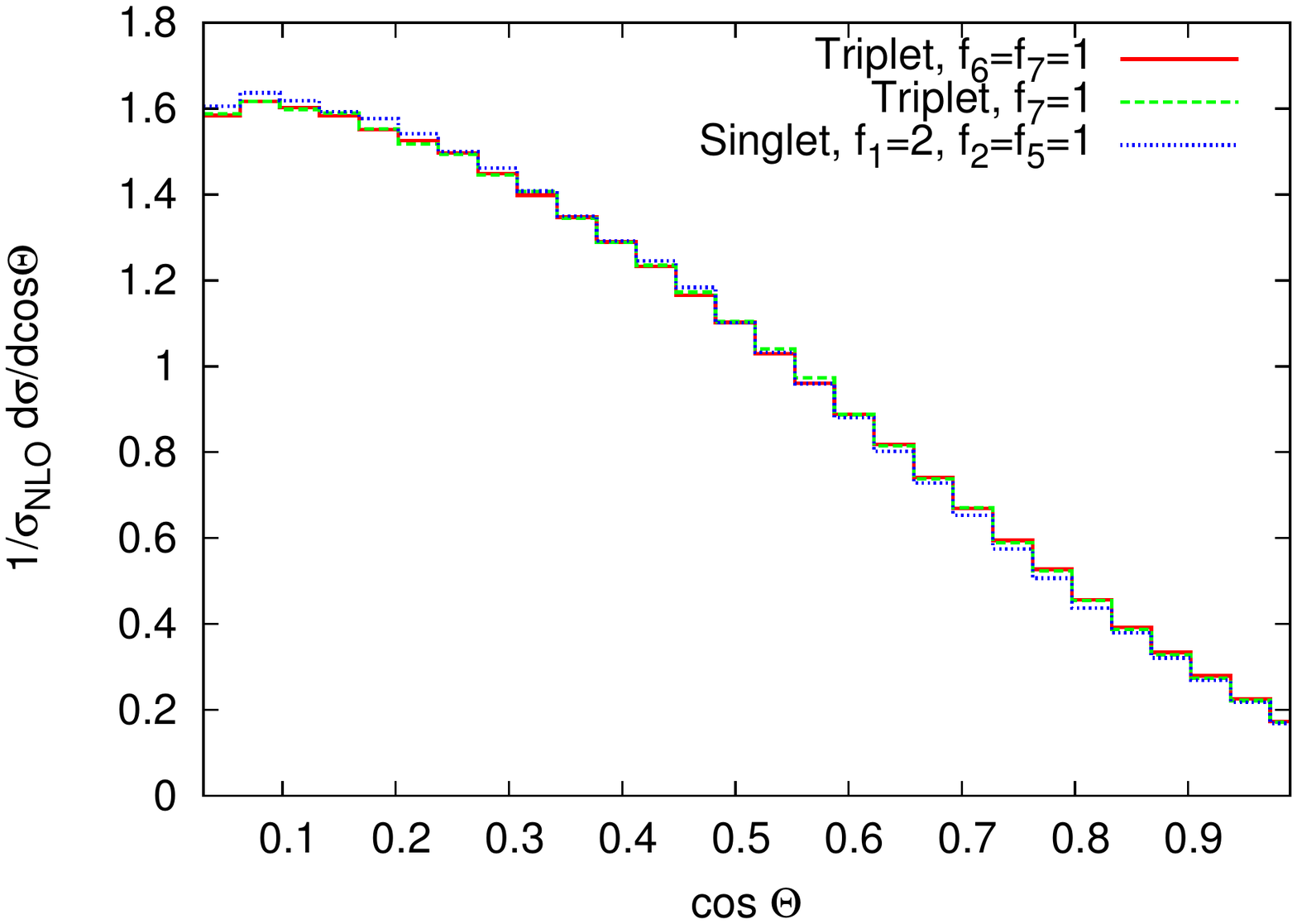}
	\end{minipage} 

\caption{Normalized $\cos \Theta$ distribution for a \spin2 resonance with different
  coupling parameters at NLO QCD accuracy. Left hand side:  \spin2 singlet,
  right hand side: \spin2 singlet and triplet.} \label{costhetav1_verschpara} 
\end{figure}

Another interesting variable is $\Theta$,  the angle between the momentum of
an initial electroweak boson and an outgoing photon in the rest frame of the
resonance. Since the dependence of the matrix element on $\Theta$ is described
by  the Wigner $d$-functions $d^j_{m,m'}(\Theta)$, even for off-shell incoming
vector bosons~\cite{HagiwaraLi}, the $\cos \Theta$ distribution should be an
indicator of the spin of the resonance. The momenta of the initial electroweak
bosons are reconstructed from those of the final-state photons and jets. In
particular, the jets are assigned to the initial quarks according to their rapidities, assuming that mainly forward scattering takes place.
Similar to $\Delta
\Phi_{jj}$, the $\cos \Theta$ distribution features a difference
between a Higgs and a \spin2 resonance (Fig.\ \ref{costhetav1}), which
is rather independent of the \spin2 couplings  (Fig.\
\ref{costhetav1_verschpara}). Therefore, it is another appropriate
distribution for a distinction between the two cases of resonances. The
NLO corrections again only shift the distributions slightly, without
modifying their characteristic shapes, which is illustrated in  Fig.\
\ref{costhetav1}.

Analogous to the $\cos \Theta$ distribution, we analyzed $\cos \Theta_{j
_{1,2}}$ distributions, where $ \Theta_{j _{1,2}}$ is the angle between
a final-state photon and the first or second tagging jet in the rest
frame of the resonance. They have characteristics similar to the $\cos
\Theta$ distribution: The distributions of a \spin2 resonance are always
significantly more central than those of the Higgs and they depend slightly on the coupling
parameters.

\begin{figure}[H]
\vspace{1.5cm}
\centerline{\includegraphics[trim=25mm 20mm 75mm 80mm, width=0.5\textwidth]{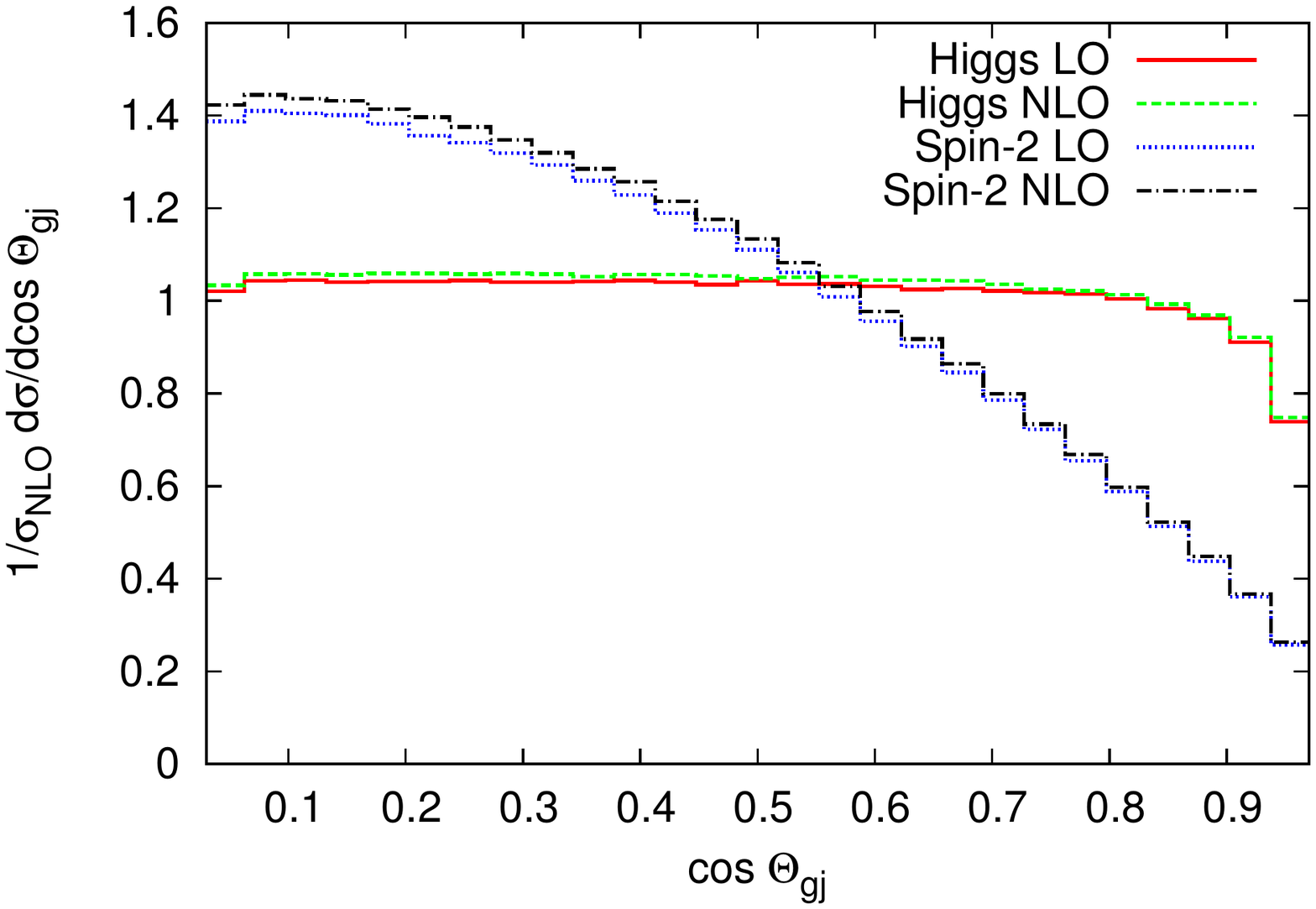}}
\caption{Distribution of the cosine of the Gottfried-Jackson angle for a Higgs and for a \spin2 resonance with couplings $f_1=0.04, f_2=0.08, f_5 = 10, f_{i\neq 1,2,5} = 0$, 
both at LO and NLO QCD accuracy.}\label{cosgja}
\end{figure}

\begin{figure}[H]
\vspace{1.5cm}
 \begin{minipage}{0.5\textwidth}%
		\includegraphics[trim=30mm 20mm 70mm 80mm, width=\textwidth]{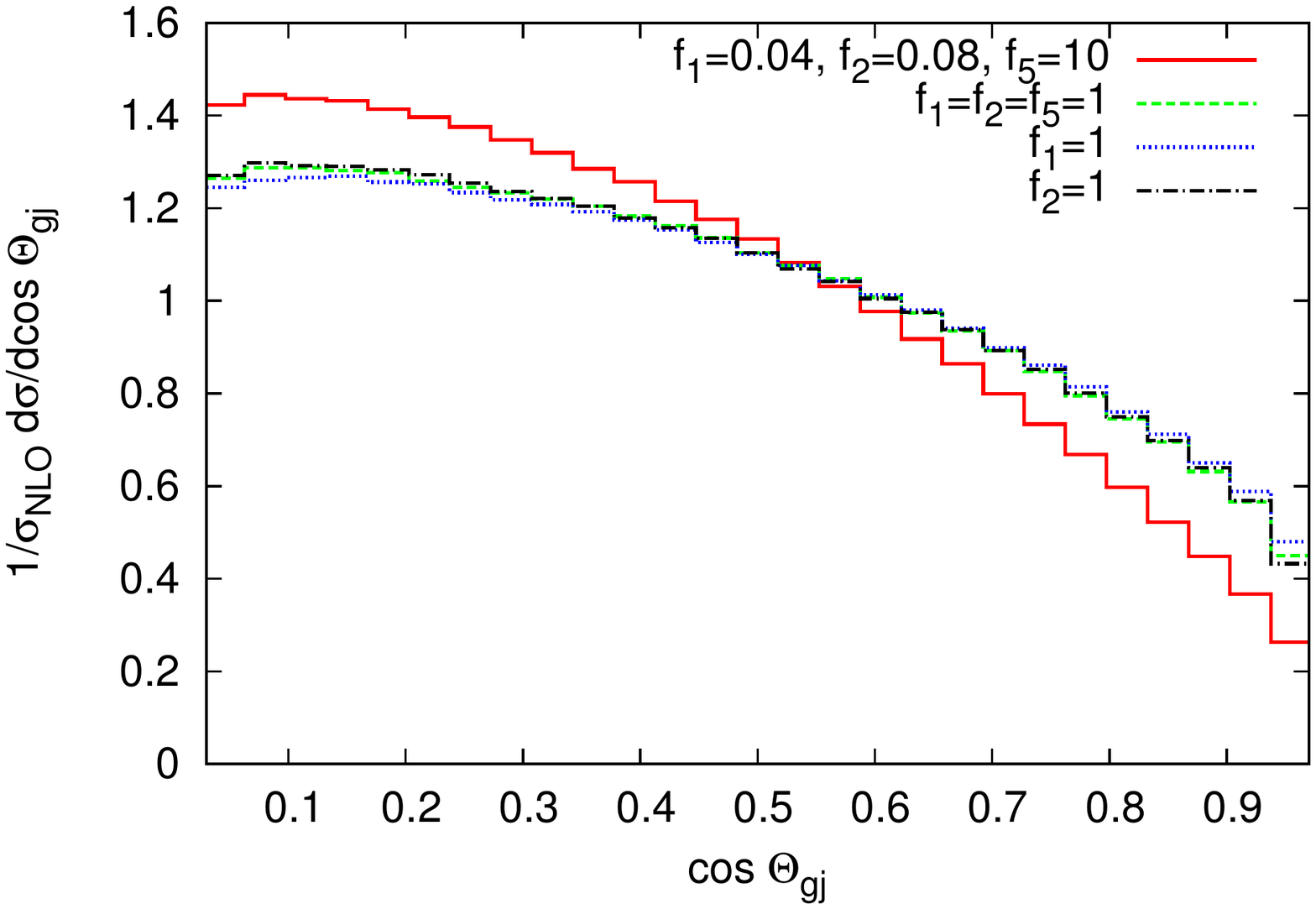}
	\end{minipage}
	\begin{minipage}{0.5\textwidth}%
		\includegraphics[trim=20mm 20mm 80mm 80mm, width=\textwidth]{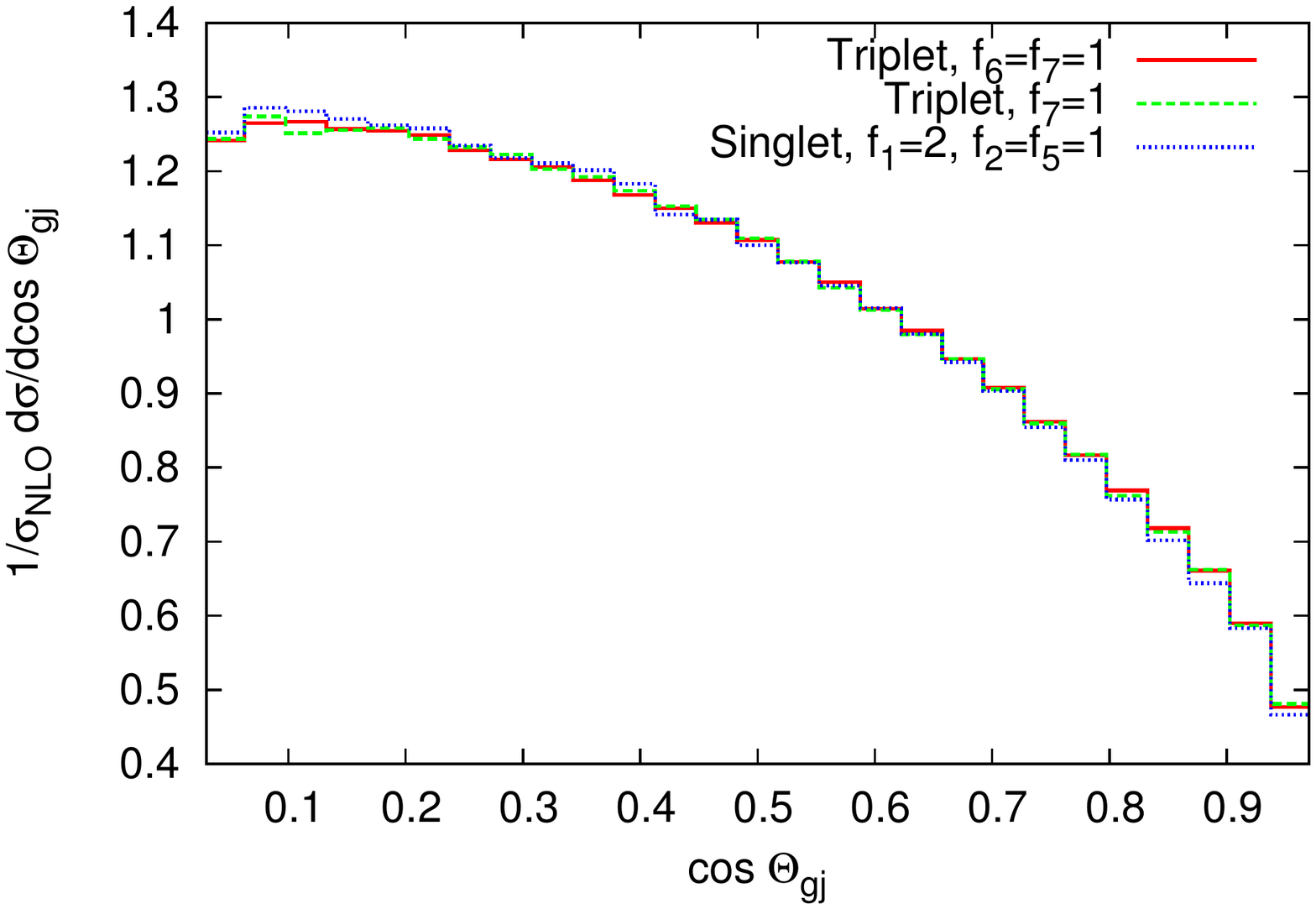}
	\end{minipage} 

\caption{Distribution of the cosine of the Gottfried-Jackson angle for a \spin2 resonance with different coupling parameters at NLO QCD accuracy. Left hand side: 
\spin2 singlet, right hand side: \spin2 singlet and triplet.} \label{cosgja_verschpara}
\end{figure}

In Fig.\ \ref{cosgja} we present another distribution which can be used for a spin-determination of the resonance: 
The cosine of the Gottfried-Jackson angle, 
which is the angle between the momentum of the \spin2 particle or the Higgs in the laboratory frame and a 
final-state photon in the rest frame of the resonance. It differs significantly between a Higgs and a \spin2 resonance and
 depends even less on the \spin2 couplings (Fig.\
\ref{cosgja_verschpara}) than the previous distributions.

\begin{figure}[H]
\vspace{1cm}
\centerline{\includegraphics[trim=25mm 20mm 50mm 55mm, width=0.5\textwidth]{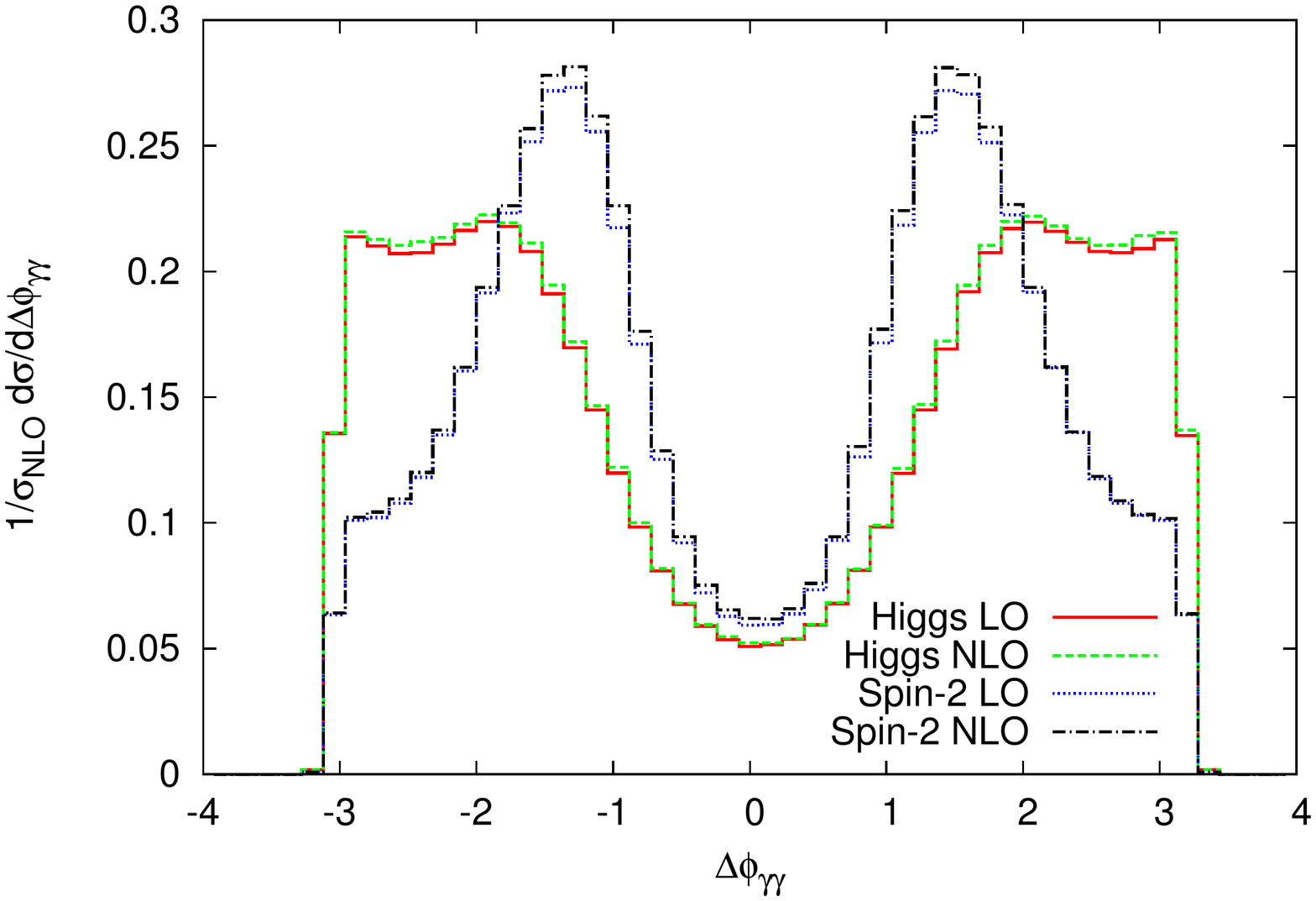}}
\caption{Azimuthal angle difference between the two photons for a Higgs
and for a \spin2 resonance with couplings $f_1=0.04, f_2=0.08, f_5 = 10, f_{i\neq 1,2,5} = 0$, 
both at LO and NLO QCD accuracy.} \label{deltaphiaa}
\end{figure}

\begin{figure}[H]
\vspace{0.5cm}
 \begin{minipage}{0.5\textwidth}%
		\includegraphics[trim=30mm 20mm 45mm 40mm, width=\textwidth]{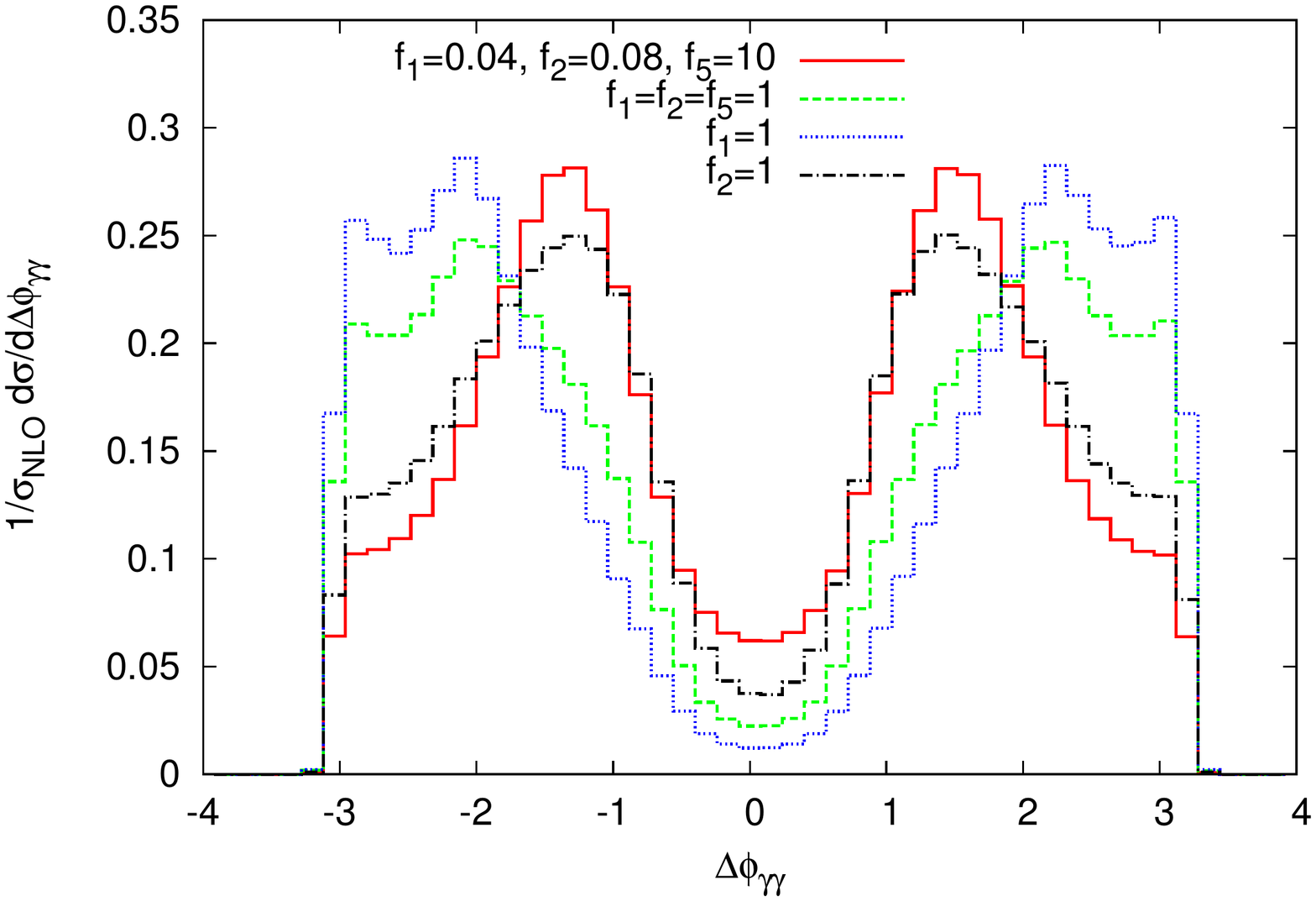}
	\end{minipage}
	\begin{minipage}{0.5\textwidth}%
		\includegraphics[trim=20mm 20mm 55mm 40mm, width=\textwidth]{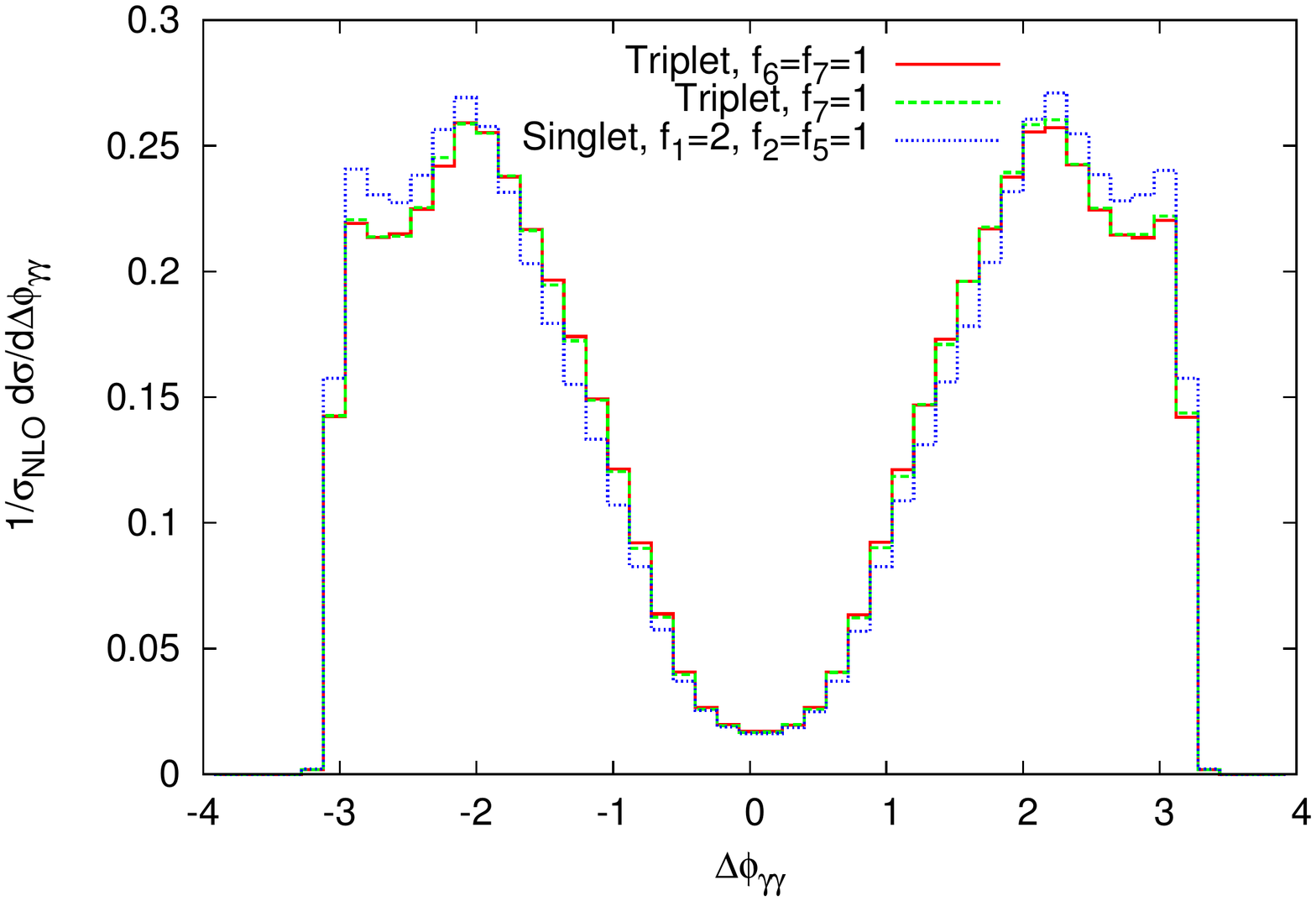}
	\end{minipage} 
\caption{Azimuthal angle difference between the two photons for a \spin2 resonance with different coupling parameters at NLO QCD accuracy. Left hand side: 
\spin2 singlet, right hand side: \spin2 singlet and triplet.} \label{deltaphiaa_verschpara}
\end{figure}

In contrast to the distributions presented before, 
the azimuthal angle difference \mbox{between} the two final-state
photons differs not only between a Higgs and a \spin2 resonance (Fig.\
 \ref{deltaphiaa}), but also between different \spin2 couplings (Fig.\
\ref{deltaphiaa_verschpara}). 
Therefore, the $\Delta \Phi_{\gamma \gamma}$
distribution is not sufficient for a spin-determination of the resonance but, 
together with other distributions, it can provide
useful information about a potential \spin2 resonance and its
parameters.

We have also investigated several other distributions, such as the rapidity distributions of the tagging jets, which are typically analyzed for VBF processes, 
yet none of them revealed any characteristic features for Higgs and \spin2 resonances.

\FloatBarrier
\subsection{Heavy Spin-2 resonances in VBF processes with four final-state leptons \label{four-lepton final state}}
\FloatBarrier

In this section, heavy \spin2 resonances in VBF processes with four leptons and two jets in the final state are investigated. Thereby, we consider the
 different leptonic final states 
 $e^+ \, e^- \, \mu^+ \mu^-$, $\, \, e^+ \, e^- \, \nu_\mu \overline{\nu}_\mu,$ $\, \, e^+ \, \nu_e \, \mu^- \, \overline{\nu}_\mu$, $ \, \, e^+ \, \nu_e \,
 \mu^+ \mu^-$ and $e^- \, \overline{\nu}_e \, \mu^+ \mu^-$. We mainly 
focus on $e^+ \, e^- \, \mu^+ \mu^-$, since a resonance in the
invariant-mass spectrum of the leptons can be reconstructed exactly if
the final state does not contain a neutrino.  We compare the 
cross sections of the different processes with and without \spin2
resonances, present the characteristic transverse-momentum and angular
distributions and study the impact of the NLO QCD corrections on these
processes. Furthermore, we investigate how the \spin2 singlet and
triplet case, as well as different coupling parameters, can be
distinguished from one another.

If not indicated otherwise, we consider a \spin2 singlet resonance with
couplings \mbox{$f_1=f_2=f_5 = 1,$} $f_{i\neq 1,2,5} = 0,\, \Lambda = 1.5\,
\text{TeV}$ and  
a mass of 1 TeV.
The parameters of the formfactor are $\Lambda_{ff} = 3 \, \text{TeV},\, n_{ff} = 4$.
For the triplet case, we use the same parameters, apart from the couplings, which are $f_6 = f_7 = 1, f_{i\neq 6,7} = 0$. 
Throughout this section, we assume a centre of mass energy of 14 TeV and we
set the mass of the Higgs boson to 130 GeV, yet the results do
not change for slightly different masses such as 126 GeV. Thus, we are
considering models with a heavy spin-2 resonance in addition to a Higgs-like
state near 126~GeV.

\begin{table}[H]
\begin{center}
\begin{tabular}{|c|c|c|c|}
\hline
Final-state leptons & Scenario & LO cross section [fb] & NLO cross section [fb] \\
\hline\hline
 & SM without \spin2 & 0.0520 & 0.0549 \\
\cline{2-4}
 $e^+ \, e^- \, \mu^+ \mu^-$ & Spin-2 singlet & 0.0541 & 0.0567 \\
\cline{2-4}
 & Spin-2 triplet & 0.0523 & 0.0557 \\
\hline\hline
 & SM without \spin2 & 0.203 &  0.212\\
\cline{2-4}
  $e^+ \, e^- \, \nu_\mu \overline{\nu}_\mu,$ & Spin-2 singlet & 0.215 &  0.226 \\
\cline{2-4}
 & Spin-2 triplet & 0.212 & 0.224 \\
\hline\hline
 & SM without \spin2 & 2.207 & 2.278 \\
\cline{2-4}
 $e^+ \, \nu_e \, \mu^- \, \overline{\nu}_\mu$ & Spin-2 singlet & 2.249 & 2.297 \\
\cline{2-4}
 & Spin-2 triplet & 2.200 & 2.267 \\
\hline\hline
 & SM without \spin2 & 0.1726 & 0.1795 \\
\cline{2-4}
 $e^+ \, \nu_e \, \mu^+ \mu^-$ & Spin-2 singlet & 0.1720 & 0.1792 \\
\cline{2-4}
 & Spin-2 triplet & 0.1734 & 0.1805\\
\hline\hline
 & SM without \spin2 & 0.0946 & 0.1001 \\
\cline{2-4}
 $e^- \, \overline{\nu}_e \, \mu^+ \mu^-$ & Spin-2 singlet & 0.0943 & 0.1000 \\
\cline{2-4}
 & Spin-2 triplet & 0.0951 & 0.1005\\
\hline
\end{tabular}
\caption{Integrated cross sections with and without \spin2 resonances for different VBF processes with four final-state 
leptons at LO and NLO QCD accuracy. 
The cuts of Section \ref{sec:settings} are applied. 
The parameter settings of the spin-2 particles can be found at the beginning 
of Section \ref{four-lepton final state}.}
 \label{crosssectionsfourlep}
\end{center}
\end{table}

Table~\ref{crosssectionsfourlep} gives a comparison of the integrated cross 
sections for the different processes with four leptons and two jets in
the final state with and without \spin2 resonances at LO and NLO QCD
accuracy.  Note that for a given process, these cross sections
correspond to a specific leptonic final state. The cross sections for
all the possible combinations of lepton generations together can be
obtained by multiplying the given cross sections with an appropriate
multiplicity factor. For some of the final states, there is some
interference between different processes, but this interference is
insignificant. One such example is $e^+ \, e^- \, \nu_e \,
\overline{\nu}_e$, which can be generated both as $(e^+ \, e^-) \,
(\nu_e \, \overline{\nu}_e)$ and as $(e^+ \nu_e) \, (e^- \, \,
\overline{\nu}_e)$, where the brackets group the fermions into pairs
connected by a continuous fermion line. The first case gives rise to
events with $m_{e^+ e^-} \approx m_Z \approx m_{\nu_e
\overline{\nu}_e}$, while the second case has $m_{e^+ \nu_e} \approx m_W
\approx m_{e^- \overline{\nu}_e}$. 

As in the photon pair-production process, the NLO corrections are relatively
small, with $K$-factors around 1.05. The statistical errors of the total cross
sections in Table~\ref{crosssectionsfourlep} are at the half percent level. 
Spin-2 resonances lead to a relative enhancement of the cross section,
which is larger for $pp \rightarrow e^+ \, e^- \, \mu^+ \mu^- \, jj$ and
$pp \rightarrow e^+ \, e^- \, \nu_\mu \overline{\nu}_\mu \, jj$ than for
the other processes, i.e.\ the relative contribution of the continuum $ZZ$
background is smaller than for the $WW$ or $WZ$ cases.  For  $pp \rightarrow
e^+ \, \nu_e \, \mu^+ \mu^- jj$ and $pp \to e^- \, \overline{\nu}_e \, \mu^+
\mu^- \,jj$, there is no singlet resonance, since only the charged resonances
of a \spin2 triplet can be produced in these processes. The \spin2 triplet
leads to a weaker enhancement than the singlet scenario throughout,
corresponding to a narrower resonance, as shown in
Table~\ref{widthsfourlep}. It should be noted, however, that the widths given
in this table are fairly arbitrary and merely reflect the parameter choices
made above. By increasing the $f_i$ by a factor of 5 or, equivalently, dropping
the scale $\Lambda$ from 1~TeV to 200~GeV, all partial and total widths and also the
\spin2 resonance contributions to the cross sections in
Table~\ref{crosssectionsfourlep} would increase by a factor of 25, making them much
more readily observable. Finally we note that for $f_1=f_2$ the decay of the
\spin2 resonance to photon pairs is as important as the decay into $ZZ$, but
the former does not suffer from the small leptonic branching ratios of $Z$
decay, which is the culprit for the small $e^+ \, e^- \, \mu^+ \mu^-$ rates 
in Table~\ref{crosssectionsfourlep}.

\FloatBarrier
\subsubsection{The process $\boldsymbol{pp \rightarrow V V \, jj \rightarrow e^+ \, e^- \, \mu^+ \mu^- \, jj}$}
\FloatBarrier

\begin{figure}
\vspace{0.5cm}
 \begin{minipage}{0.5\textwidth}%
		\includegraphics[trim=40mm 20mm 15mm 35mm, width=\textwidth]{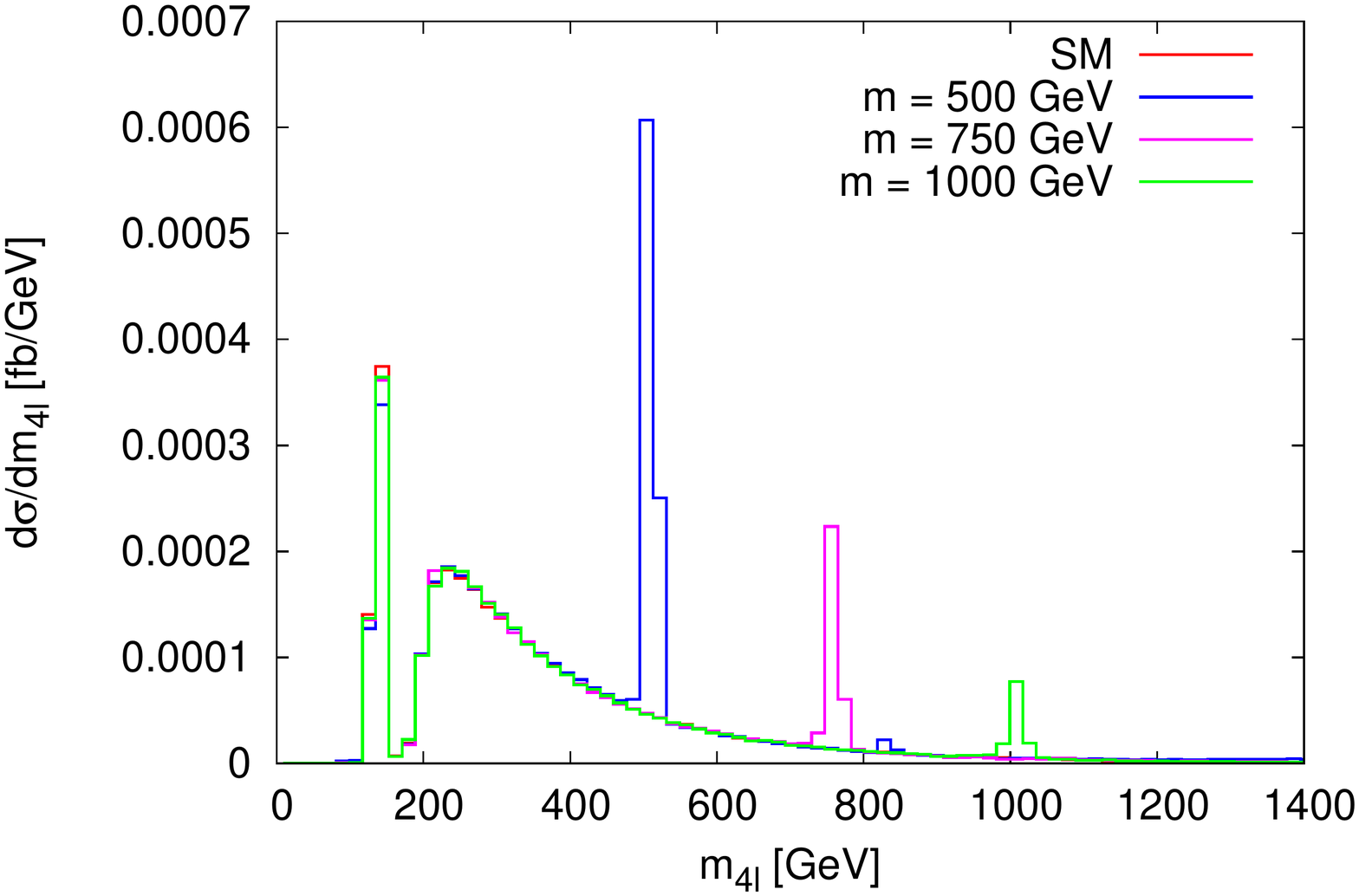}
	\end{minipage}
	\begin{minipage}{0.5\textwidth}%
		\includegraphics[trim=30mm 20mm 25mm 35mm, width=\textwidth]{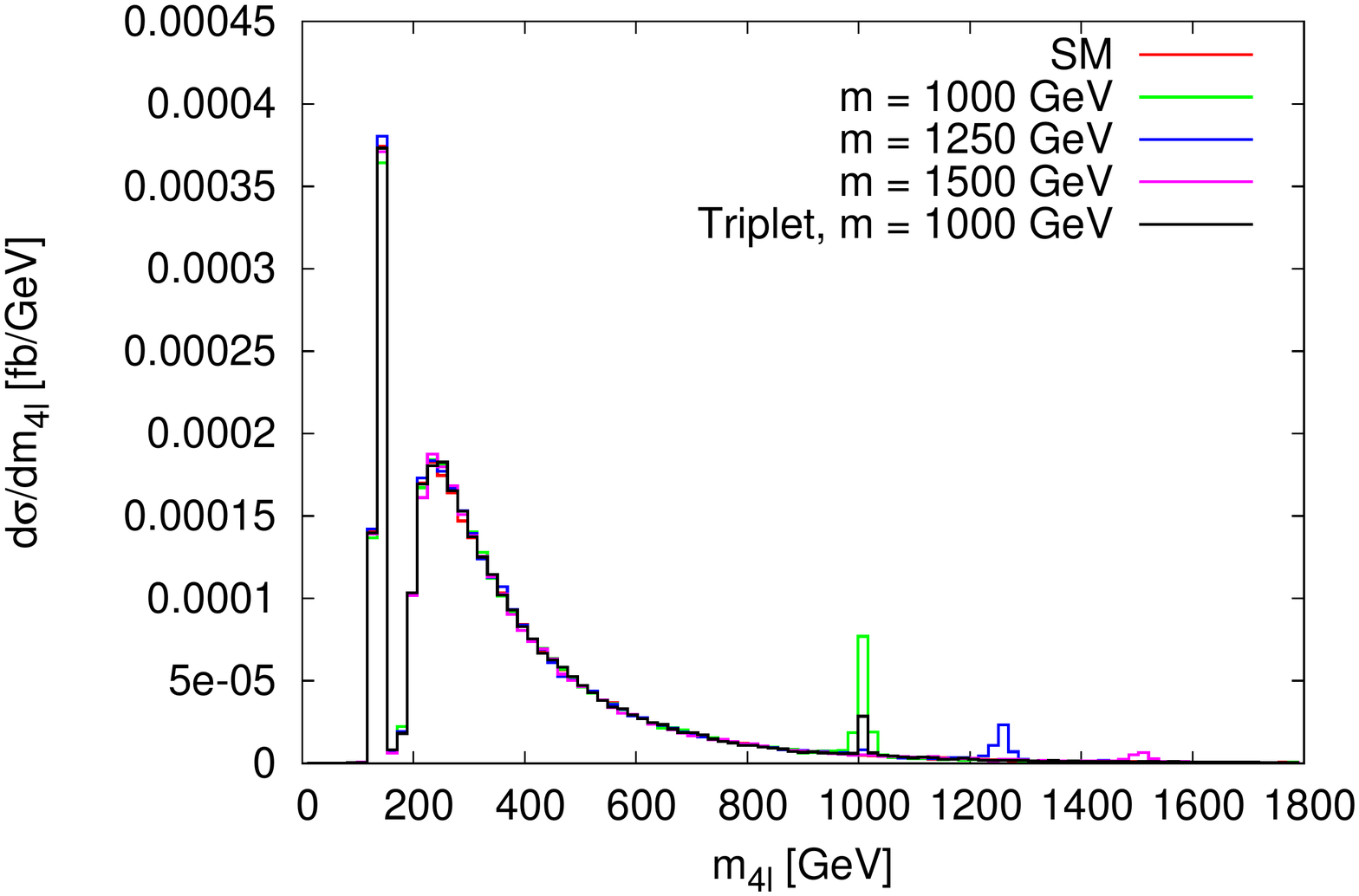}
	\end{minipage} 
\vspace{-0.2cm}
\caption{Invariant-mass distribution of the four final-state leptons: Spin-2 singlet and triplet resonances with different masses in the process 
\mbox{$pp \rightarrow V V \, j j \rightarrow e^+ \, e^- \, \mu^+ \mu^- \, jj$} at NLO QCD accuracy.} \label{mlep210}
\end{figure}

Fig.~\ref{mlep210} depicts the invariant-mass distribution of the four
final-state leptons for the process ${pp \rightarrow V V \, jj
\rightarrow e^+ \, e^- \, \mu^+ \mu^- \, jj}$, which can be fully
reconstructed experimentally since there are no neutrinos in the final
state. For the Standard Model, a Higgs resonance at 126~GeV is followed
by a continuous distribution which vanishes for high energies.  The
\spin2 singlet resonance peak is shown for different masses up to 1.5 TeV for
the given parameter settings. The triplet case is analogous, except for the
height and width of the resonance, which is chosen to have a mass of 1 TeV. The
triplet resonance is generated by the neutral \spin2 triplet particle in this
process.  Due to the formfactor, there are no unphysical high-energy
contributions outside the depicted mass range, which otherwise would
result from unitarity violation. For a mass of 500 GeV, these
contributions are not suppressed completely for $\Lambda_{ff} = 3 \,
\text{TeV}$. For such small masses, a smaller value of $\Lambda_{ff}$
should be chosen. The total widths of the \spin2 resonances shown in
Fig.~\ref{mlep210} are given in Table~\ref{widthsfourlep}.

\begin{table}
\vspace{0.5cm}
\begin{center}
\begin{tabular}{|l|c|}
\hline
\hspace{0.5cm} Resonance & Width [GeV]\\
\hline\hline
Singlet, 500 GeV& 0.982\\
\hline
Singlet, 750 GeV& 3.238\\
\hline
Singlet, 1000 GeV& 7.607\\
\hline
Singlet, 1250 GeV& 14.795\\
\hline
Singlet, 1500 GeV& 25.505\\	
\hline
Triplet, 1000 GeV& 1.004\\
\hline
\end{tabular}
\caption{Total widths of the \spin2 resonances shown in Fig.~\ref{mlep210}.} \label{widthsfourlep}
\end{center}
\end{table}

\begin{figure}
\vspace{1cm}
 \begin{minipage}{0.5\textwidth}%
		\includegraphics[trim=30mm 20mm 45mm 55mm, width=\textwidth]{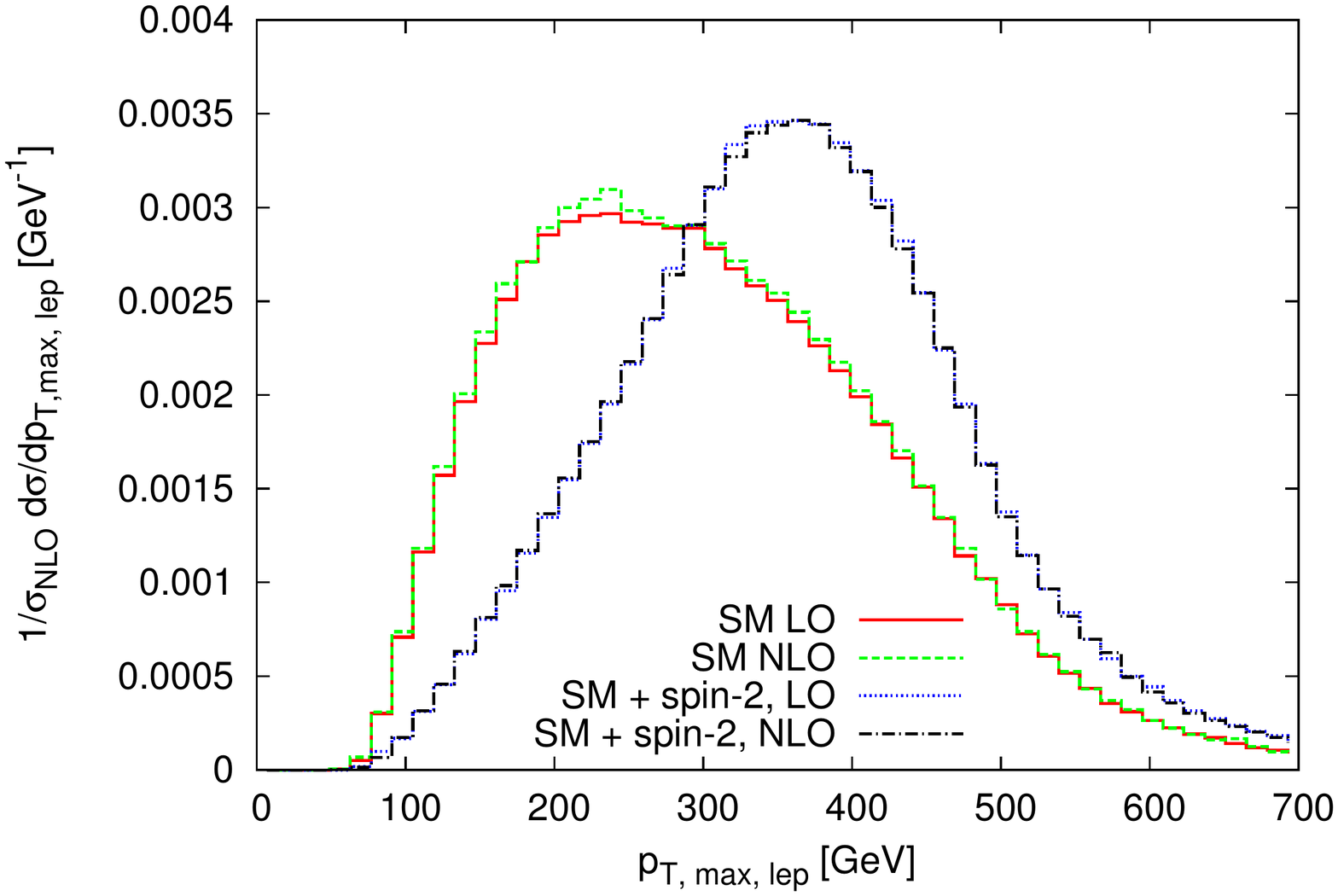}
	\end{minipage}
	\begin{minipage}{0.5\textwidth}%
		\includegraphics[trim=20mm 20mm 55mm 55mm, width=\textwidth]{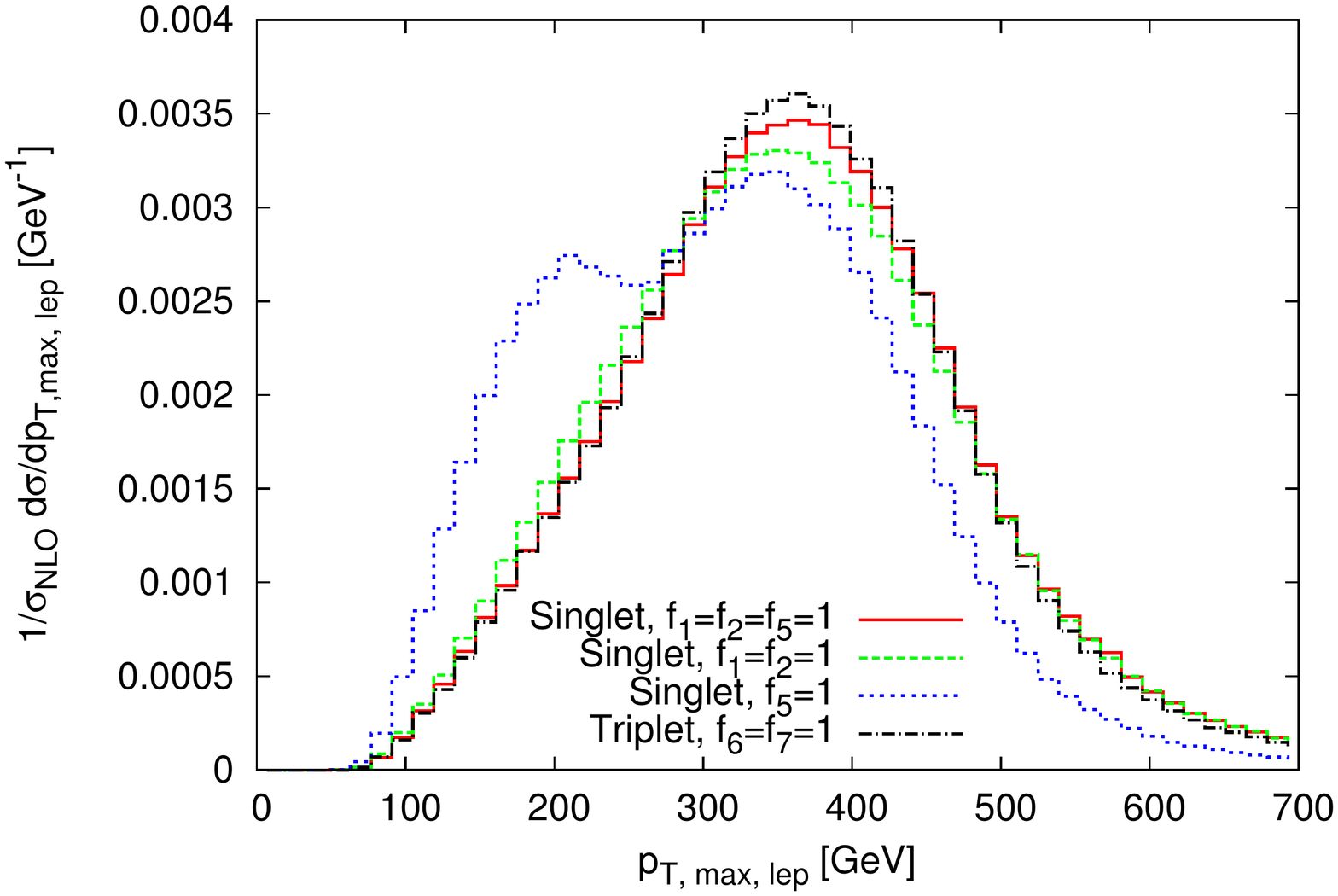}
	\end{minipage} 
\vspace{-0.2cm}
\caption{Transverse-momentum distribution of the hardest final-state lepton for 
events near the spin-2 resonance (see text for details). 
Left hand side: With and without a \spin2 singlet resonance with couplings 
$f_1=f_2=f_5 = 1$ at LO and NLO QCD accuracy. 
 Right hand side: Spin-2 singlet and triplet resonance with different 
coupling parameters at NLO accuracy. 
The electroweak continuum contributions as mentioned in Section 
\ref{sec:Calculation} are always included.} \label{ptmaxlep210}
\end{figure}

\begin{figure}
\vspace{1.0cm}
 \begin{minipage}{0.5\textwidth}%
		\includegraphics[trim=30mm 20mm 70mm 80mm, width=\textwidth]{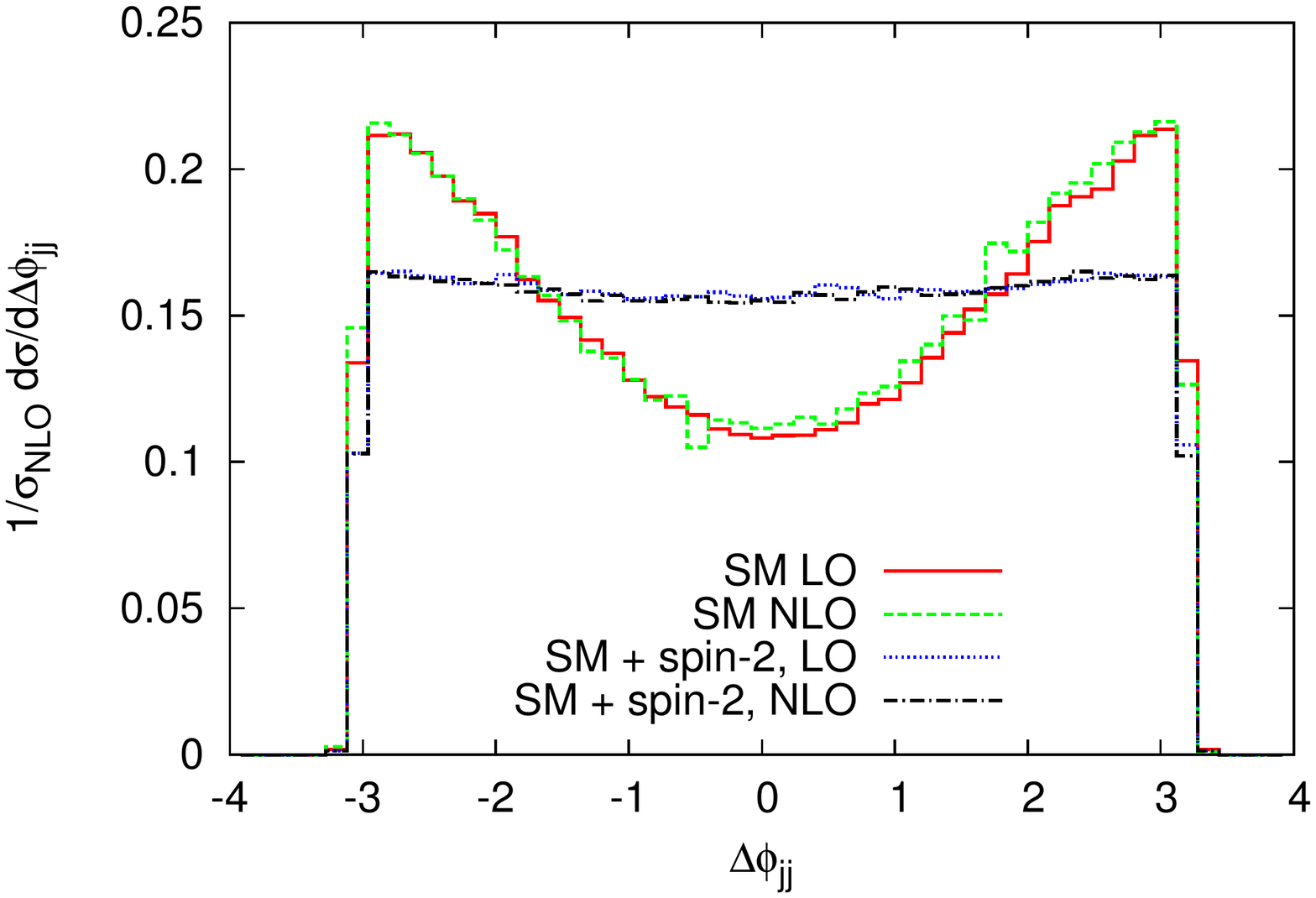}
	\end{minipage}
	\begin{minipage}{0.5\textwidth}%
		\includegraphics[trim=20mm 20mm 80mm 80mm, width=\textwidth]{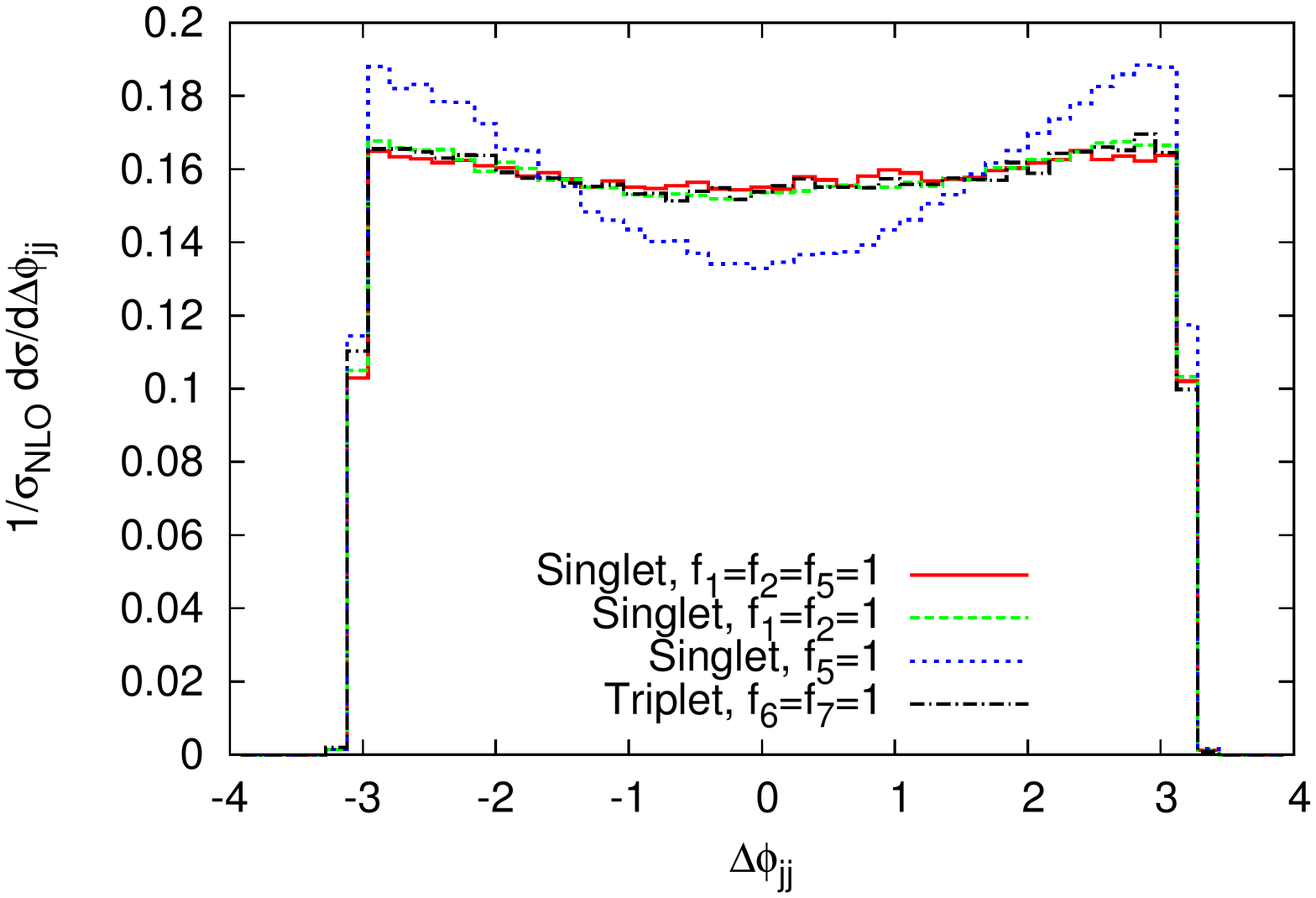}
	\end{minipage} 
\vspace{-0.2cm}
\caption{Azimuthal angle difference between the two tagging jets. Parameters for the 
different cases are as in Fig.~\ref{ptmaxlep210}.} \label{deltaphijets210}
\end{figure}

\begin{figure}
\vspace{1.0cm}
 \begin{minipage}{0.5\textwidth}%
		\includegraphics[trim=30mm 20mm 70mm 80mm, width=\textwidth]{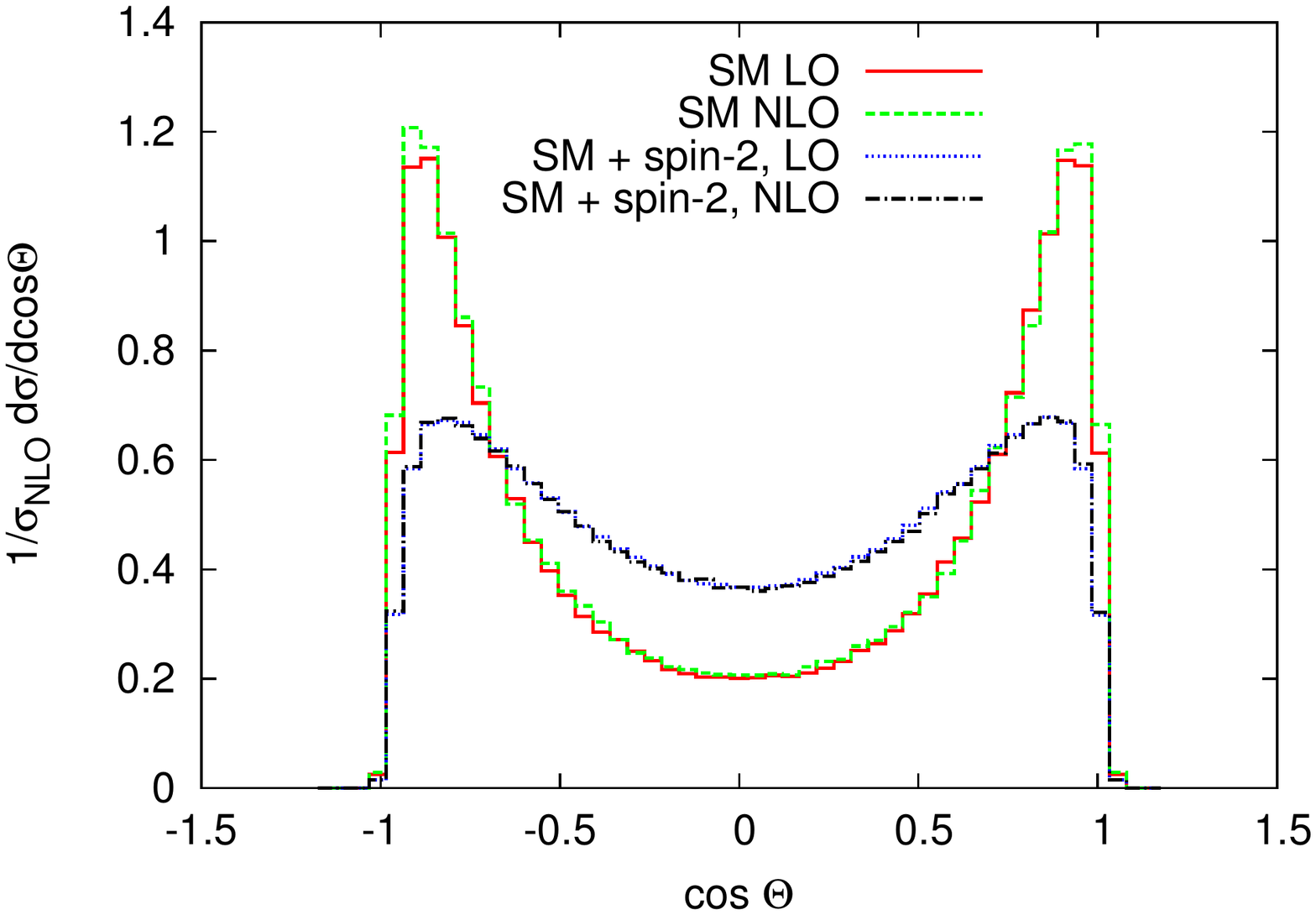}
	\end{minipage}
	\begin{minipage}{0.5\textwidth}%
		\includegraphics[trim=20mm 20mm 80mm 80mm, width=\textwidth]{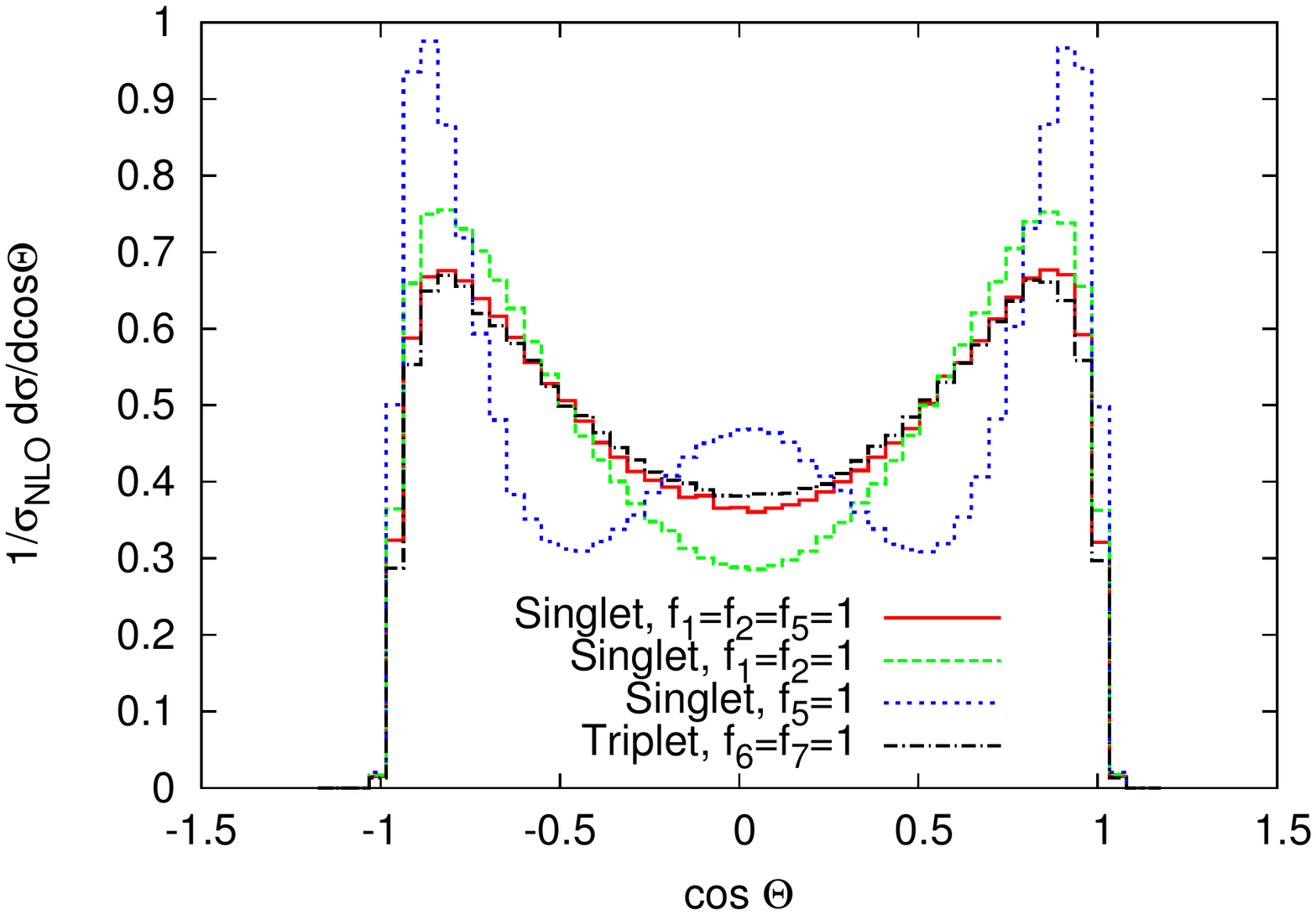}
	\end{minipage} 
\vspace{-0.2cm}
\caption{Cosine of the angle between the momenta of an incoming and an
outgoing electroweak boson in the rest frame of the \spin2 resonance. 
Parameters for the different cases are as in Fig.~\ref{ptmaxlep210}.} 
\label{costhetav1210}
\end{figure}

\begin{figure}
\vspace{1.0cm}
 \begin{minipage}{0.5\textwidth}%
		\includegraphics[trim=30mm 20mm 70mm 80mm, width=\textwidth]{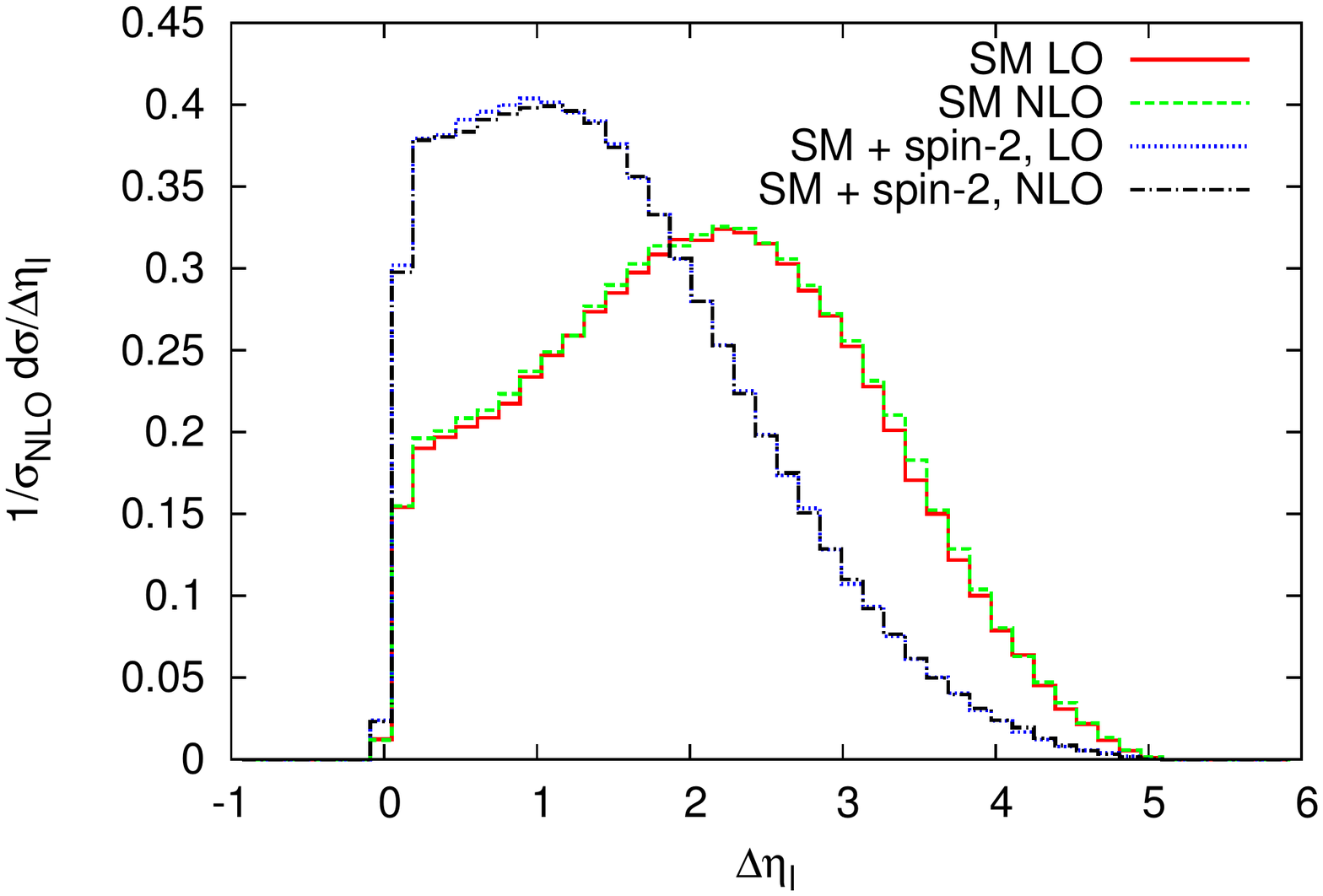}
	\end{minipage}
	\begin{minipage}{0.5\textwidth}%
		\includegraphics[trim=20mm 20mm 80mm 80mm, width=\textwidth]{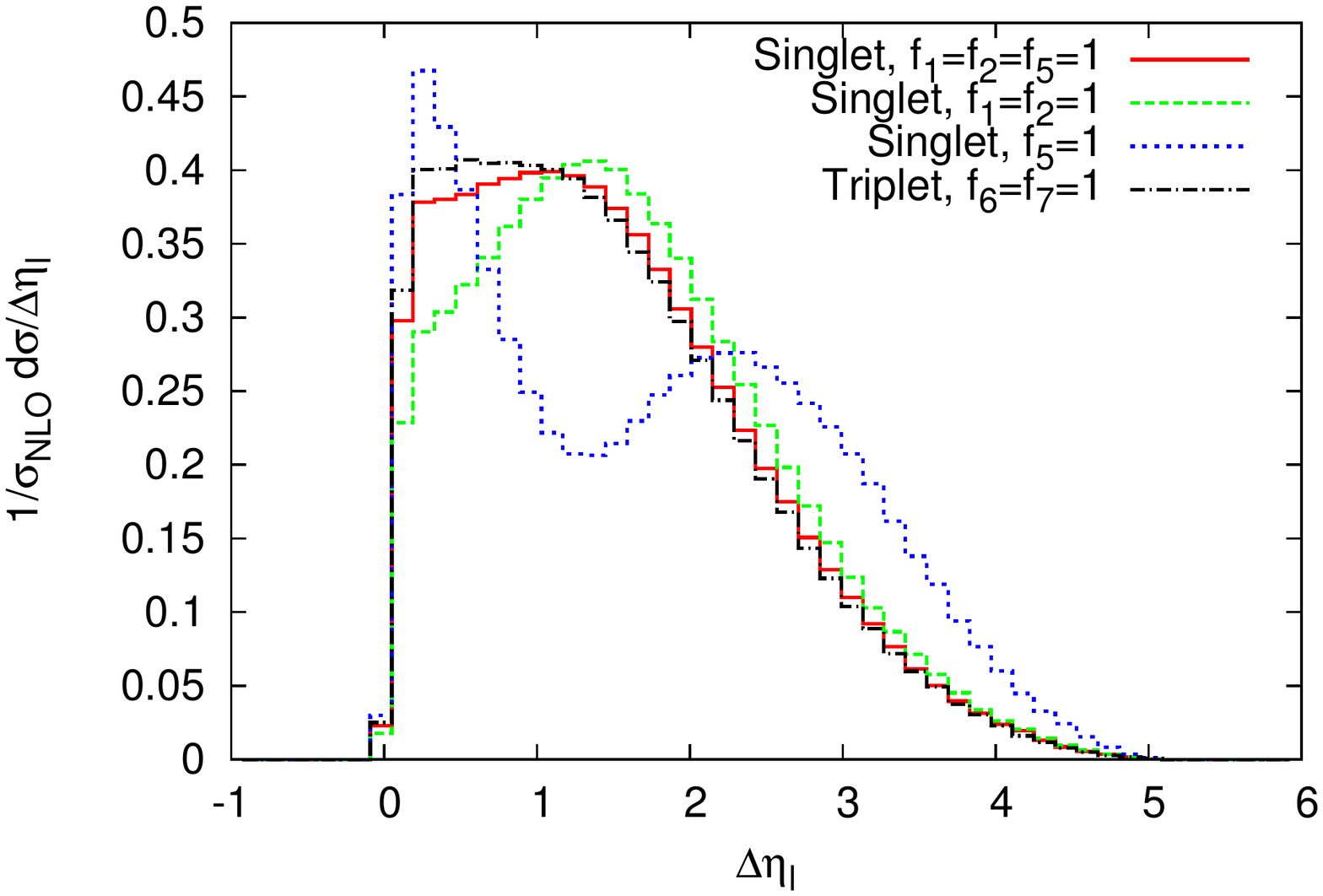}
	\end{minipage} 
\vspace{-0.2cm}
\caption{Pseudorapidity difference between the two positively charged final-state leptons.
Parameters for the different cases are as in Fig.~\ref{ptmaxlep210}.} \label{deltaetalep210}
\end{figure}

In Figs.~\ref{ptmaxlep210} to~\ref{deltaetalep210}, we present the
characteristic transverse-momentum and angular distributions of \spin2 singlet and 
triplet resonances at 1 TeV. We have selected those distributions which
show the most prominent differences between the different models.
The left hand sides depict the distributions for a \spin2 singlet resonance 
(including the electroweak continuum in a mass bin around the resonance) 
and of the SM continuum expectation, without such a resonance, at LO and NLO QCD accuracy, 
whereas the right hand sides compare the singlet and triplet case and different 
coupling parameters at NLO QCD accuracy, again including the electroweak 
continuum in all cases. All the figures are normalized to 
the NLO cross sections. In order to reveal the features of the \spin2 
resonances, additional cuts on the invariant mass of the four leptons are applied. 
For the parameter settings $f_1=f_2=f_5 = 1$, $f_1=f_2 = 1$ and for the
SM without a \spin2 resonance, they are chosen as $m_{4l}=1000 \pm 50$
GeV. For the triplet case, we use  $m_{4l}=1000 \pm 10$ GeV and for $f_5
= 1$, we set $m_{4l}=1000 \pm 5$ GeV. The latter cases are only
presented for illustration, since the experimental resolution is expected to be
worse. However, for larger couplings $f_i/\Lambda$ and resulting larger
production rates of the \spin2 resonances, the characteristic distributions
would be visible for wider mass bins. With these additional cuts, we
obtain a signal-to-background ratio of approximately one for the case
$f_5=1$, approximately three for $f_1=f_2 = 1$ and approximately four in the other cases.
Here ``background'' refers to the expected electroweak continuum contribution 
from VBF within the SM.

Characteristic differences between a \spin2 resonance and the SM background arise especially in the distribution of the transverse momentum of the hardest 
final-state lepton (Fig.\ \ref{ptmaxlep210}), the azimuthal angle difference between the two tagging jets (Fig.\ \ref{deltaphijets210}), the cosine of the angle 
between the momenta of an incoming 
and an outgoing electroweak boson in the rest frame of the \spin2 resonance (or of the four final-state leptons, respectively) 
(Fig.\ \ref{costhetav1210}) and the pseudorapidity difference between the two 
positively charged final-state leptons (Fig.\ \ref{deltaetalep210}). 
The NLO corrections do not have any significant impact on cross sections and 
distributions in the high invariant-mass region studied here. 
A \spin2 triplet resonance resembles a singlet resonance with couplings 
$f_1=f_2=f_5 = 1$. The cases $f_1=f_2=f_5 = 1$ and $f_1=f_2 = 1$ are difficult to 
distinguish since, numerically, the $f_5$ contribution is sub-dominant.  
However, small differences are visible in the $\cos \Theta$ and $\Delta \eta_l$ 
distributions and they do not only stem from contamination of the electroweak 
continuum. The coupling $f_5$ alone leads to different distributions throughout. 
This is not simply an effect of the sizable electroweak continuum background for small
$f_5/\Lambda$, 
but reflects the different tensor structure, as we have verified by comparing 
with the case $f_5=10$. For $f_5=10$, the $\Delta\Phi_{jj}$ distribution
 approaches those of the other \spin2 cases, while for the $\cos \Theta$ distribution 
the peak around $\cos \Theta = 0$ becomes more prominent.

Note that the $\cos \Theta$ distribution of Fig.~\ref{costhetav1210} is not accessible 
experimentally for processes with final-state neutrinos. However, since 
for this figure the momenta of the electroweak bosons were not reconstructed 
from final-state momenta, as in Section \ref{sec:photonpairproduction}, but taken 
directly from the Monte Carlo information, the results can be directly taken over for 
final states with neutrinos. 
Besides $\cos \Theta$, we have also studied the $\cos \Theta_{j_{1,2}}$ 
distributions, where $ \Theta_{j _{1,2}}$ is the angle between
the momenta of an outgoing electroweak boson and the first or second
tagging jet in the rest frame of the \spin2 resonance (or of the four
final-state leptons, respectively). These 
distributions show a behaviour similar to $\cos \Theta$.

\FloatBarrier

 \subsubsection{Other processes with four final-state leptons}
The VBF process $pp \rightarrow V V \, jj \rightarrow e^+ \, e^- \, \nu_\mu \overline{\nu}_\mu \, jj$ is very similar to 
$pp \rightarrow V V \, jj \rightarrow e^+ \, e^- \, \mu^+ \mu^- \, jj$,
which was analyzed in the previous section. Theoretically, a \spin2 resonance 
in the invariant four-lepton mass spectrum as 
well as the transverse-momentum and angular distributions with a cut on the invariant four-lepton mass look the same, despite the fact that there is no use in 
investigating the angular distributions of the two charged leptons, since they emerge from the same vector boson in this case. However, since the invariant 
four-lepton mass cannot be reconstructed experimentally, the transverse mass of the lepton system $e^+ \, e^- \, \nu_\mu \overline{\nu}_\mu$ has to be 
considered instead, which is defined as~\cite{Englert:2008tn}:
\begin{equation}
 m_T=\sqrt{(E_{T,ll}+E_{T,\text{miss}})^2-(\boldsymbol{p}_{T,ll}+\boldsymbol{p}_{T,\text{miss}})^2}, \label{mtrans}
\end{equation}
with
\begin{equation}
E_{T,ll}=\sqrt{\boldsymbol{p}_{T,ll}^2+m_Z^2} \hspace{0.2cm},  \hspace{0.5cm} E_{T,\text {miss}}=\sqrt{\boldsymbol{p}_{T,\text{miss}}^2+m_Z^2},
\end{equation}
where $E_{T,ll}, \boldsymbol{p}_{T,ll}$ denote the transverse energy and momentum of the two charged leptons and $E_{T,\text{miss}}, 
\boldsymbol{p}_{T,\text{miss}}$ those of the neutrino system.

Although an excess from the \spin2 resonance is hardly visible in the transverse-mass spectrum for our choice of parameters, some of the characteristics of the 
differential distributions remain accessible if a transverse-mass cut, e.g. $m_T=1000 \pm 100$ GeV, is applied instead of the cut on the invariant four-lepton 
mass. Whereas the difference in the transverse momentum 
of the hardest lepton (left hand side of Fig.\ \ref{ptmaxlep210}) disappears, the azimuthal angle difference of the two tagging jets (Fig.\ \ref{deltaphijets210}) 
remains the same. 

\begin{figure}
\vspace{0.5cm}
 \begin{minipage}{0.5\textwidth}%
		\includegraphics[trim=40mm 20mm 15mm 35mm, width=\textwidth]{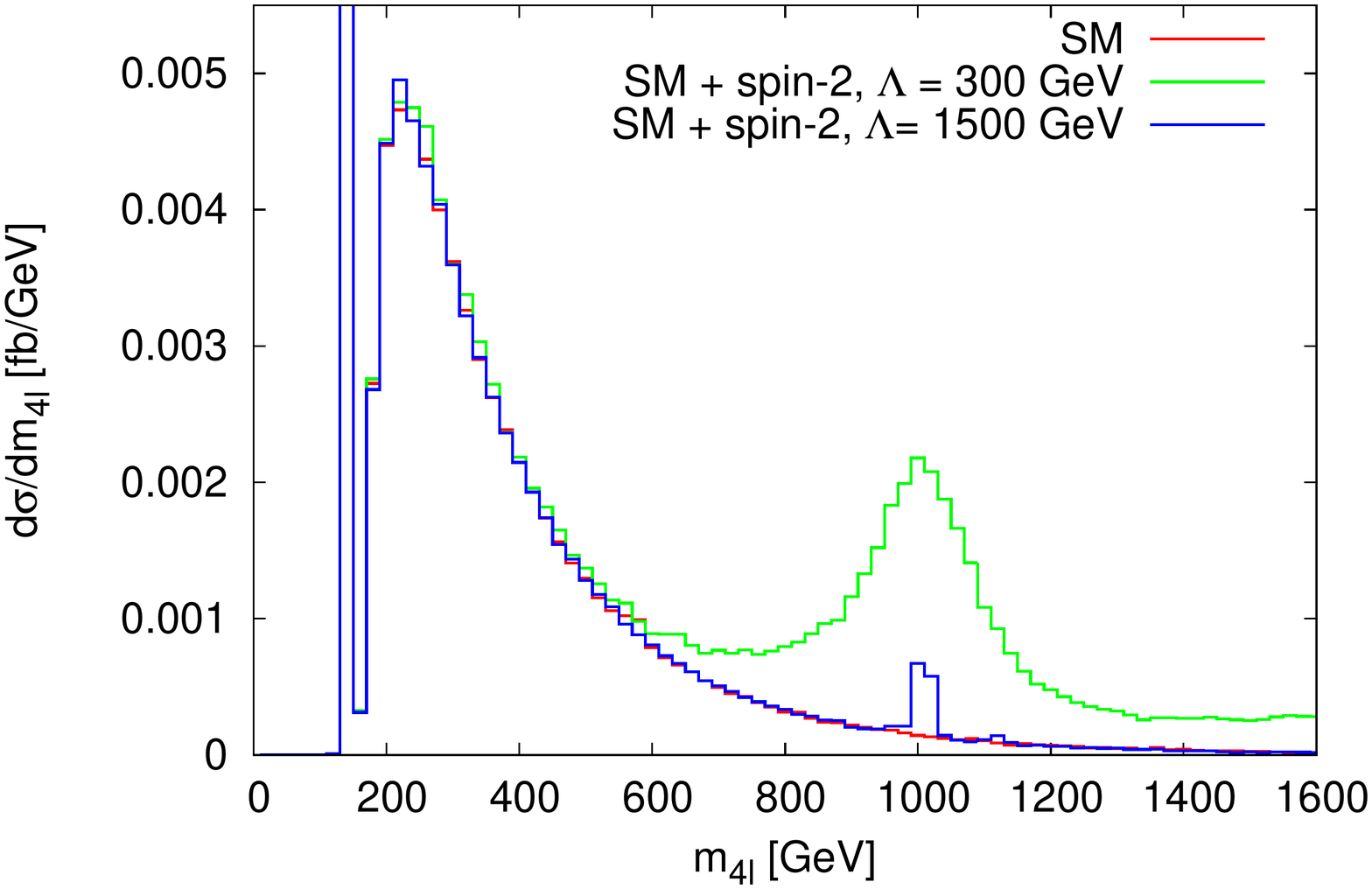}
	\end{minipage}
	\begin{minipage}{0.5\textwidth}%
		\includegraphics[trim=30mm 20mm 25mm 35mm, width=\textwidth]{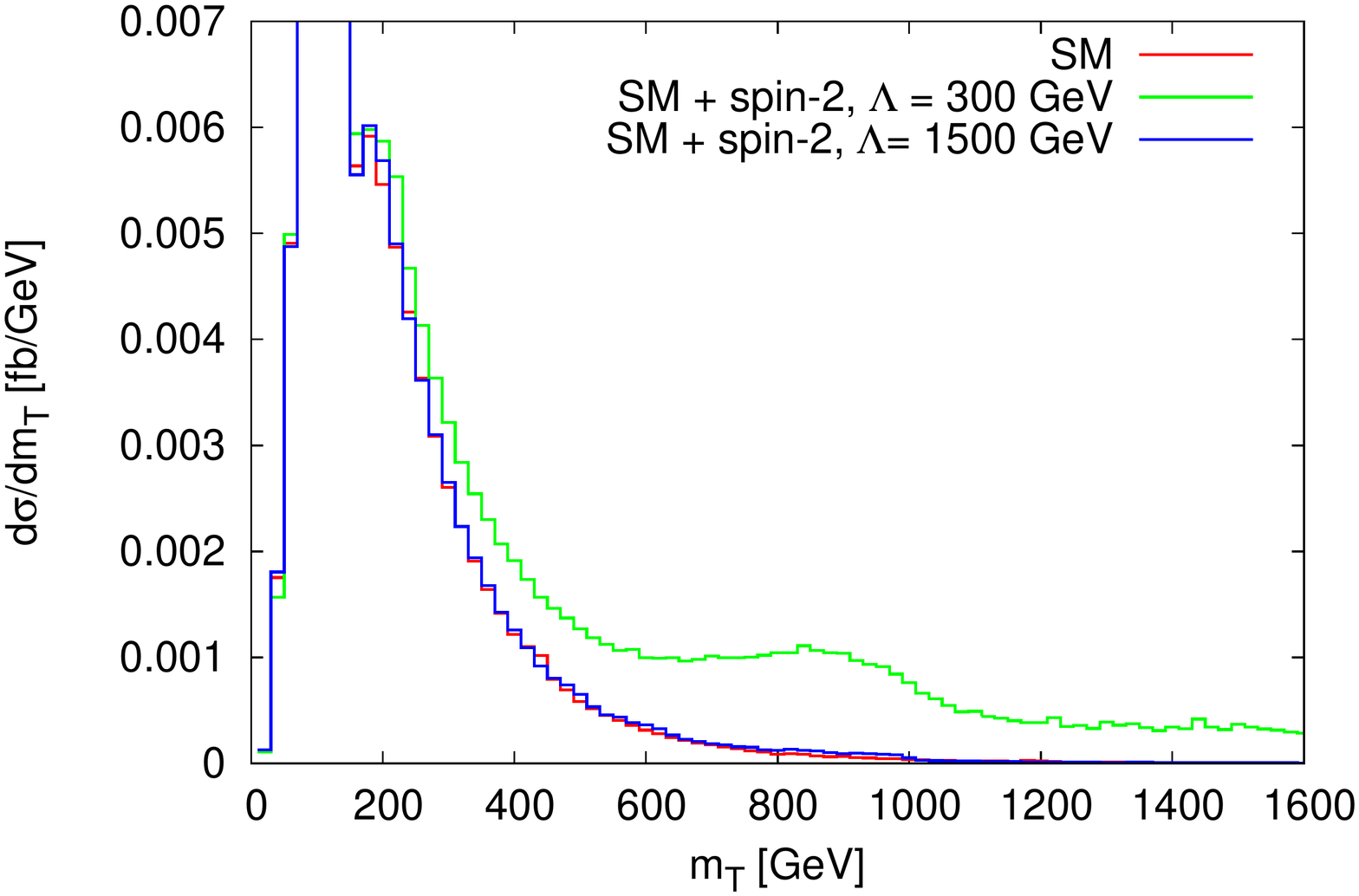}
	\end{minipage} 
\vspace{-0.2cm}
\caption{Process \mbox{$pp \rightarrow W^+ W^- \, jj \rightarrow e^+ \,
\nu_e \, \mu^- \, \overline{\nu}_\mu \, jj$} with and without a \spin2 
singlet resonance for different values of $\Lambda$. Left hand side: Invariant mass
of the four final-state leptons, right hand side: Transverse mass, at NLO QCD accuracy.} \label{mlep200}
\end{figure}

In the process $pp \rightarrow W^+ W^- \, jj \rightarrow e^+ \, \nu_e \,
\mu^- \, \overline{\nu}_\mu \, jj$, it is even harder to access the
characteristics of \spin2 resonances, since \mbox{$t \bar{t} + \text{jets}$} constitute a
large background to this process at the LHC.  Furthermore, the \spin2
singlet resonance is smaller than in the processes analyzed before (see
Fig.~\ref{mlep200}) and the triplet resonance is even smaller,
since the uncharged triplet particle couples to two $W$ bosons only via
the $f_6$ term, whereas the Feynman rules for vertices involving photons
and Z bosons contain the coupling $f_7$ (see Eq.
\ref{tripletFeynmanrules}).  Again, the invariant four-lepton mass is
not accessible experimentally and the transverse mass of the system $e^+
\, \nu_e \, \mu^- \, \overline{\nu}_\mu$ has to be considered instead.
For this process, it is defined as in Eq. \ref{mtrans}, but with
\cite{Jager:2006zc}
\begin{equation}
E_{T,ll}=\sqrt{\boldsymbol{p}_{T,ll}^2+m_{ll}^2} \hspace{0.2cm},  \hspace{0.5cm}
E_{T,\text {miss}}=\sqrt{\boldsymbol{p}_{T,\text{miss}}^2+m_{\nu \nu}^2}  
\rightarrow |\boldsymbol{p}_{T,\text{miss}}|,
\end{equation}
where $m_{ll}$ is the invariant mass of the charged-lepton system.

Fig.~\ref{mlep200} compares \spin2 singlet resonances for different
values of $\Lambda$ in the (only theoretically accessible) four-lepton
invariant mass and the transverse-mass distribution. Here, the high
Higgs resonance peak is cut off in order to concentrate on the \spin2
resonance region.  For the usual parameter settings, with $\Lambda=1.5$
TeV, the transverse-mass spectrum is roughly the same for the SM with
and without a \spin2 resonance. Even for $\Lambda$ as small as 300 GeV
(or large couplings $f_i$, respectively), the resonance is very smeared out. 
The characteristics of the transverse-momentum and angular
distributions, which are theoretically similar to those of the process
$pp \rightarrow V V \, jj \rightarrow e^+ \, e^- \, \mu^+ \mu^- \, jj$,
remain accessible with a transverse-mass cut if the couplings are not
too small. For the usual settings, with $\Lambda=1.5$ TeV, the
differences between the SM and a model with a \spin2 singlet resonance
are small and very difficult to observe in the $W^+W^-$ mode at the LHC.

In the processes $pp \rightarrow V V \, jj \rightarrow e^+ \, \nu_e \, \mu^+ \mu^- jj$ and $pp \rightarrow V V \, jj \rightarrow e^- \, \overline{\nu}_e \, \mu^+ \mu^- \, jj$, 
only charged resonances are possible. Therefore, there can be a \spin2
triplet resonance generated by the charged triplet particle, but no
singlet resonance.  Again, this resonance can only be reconstructed in
Monte Carlo in the invariant four-lepton mass spectrum and the features of the distributions with a cut on
the invariant four-lepton mass are as usual. The corresponding
transverse mass in this case reads \cite{Englert:2008tn}:
\begin{equation}
 m_T=\sqrt{(E_{T,lll}+E_{T,\text{miss}})^2-(\boldsymbol{p}_{T,lll}+\boldsymbol{p}_{T,\text{miss}})^2}, \label{mtrans_220}
\end{equation}
with
\begin{equation}
E_{T,lll}=\sqrt{\boldsymbol{p}_{T,lll}^2+m_{lll}^2} \hspace{0.2cm},  \hspace{0.5cm} E_{T,\text {miss}}= |\boldsymbol{p}_{T,\text{miss}}|,
\end{equation}
where $m_{lll}$ denotes the invariant mass of the charged-lepton system, $E_{T,lll}, \boldsymbol{p}_{T,lll}$ its transverse energy and momentum and $E_{T,\text{miss}}, 
\boldsymbol{p}_{T,\text{miss}}$ those of the neutrino.\\
The \spin2 triplet resonance peak can be observed in the transverse-mass spectrum 
if the couplings are not too small. However, we find that the parameters chosen 
above yield a marginal signal only.
With a transverse-mass cut of $m_T=1000 \pm 100$ GeV, the characteristics of 
distributions, such as the pseudorapidity difference between two  
final-state leptons of the same charge, can be studied and yield results 
similar to those found for final states with four charged leptons.

\FloatBarrier\section{Conclusions \label{sec:conclusions}}
We have studied \spin2 resonances in vector-boson-fusion processes at
the LHC within the framework of an effective model describing the
interaction of  \spin2 particles with electroweak gauge bosons for a
\spin2 singlet and a \spin2 triplet scenario. Unitarity of the underlying
vector boson scattering in such models requires the introduction of
formfactors which decrease the contribution of \spin2 particle exchange at
high energies and which parametrize
high-energy contributions beyond this effective model.
The calculation was
performed within the Monte Carlo program \textsc{Vbfnlo} at NLO QCD
accuracy. We have analyzed two different kinds of processes: light,
Higgs-like \spin2 resonances producing two photons and heavy \spin2
resonances in different four-lepton final states. 

For the first process, by using formfactors and
adjusting the couplings it is possible to tune the cross section and the
transverse-momentum distributions of photons and jets to roughly agree  with
expectations for the SM Higgs boson of the same mass. Even then, the
azimuthal-angle difference between the two tagging jets, as well as the
Gottfried-Jackson angle and the angle between an intermediate vector boson and
a final-state photon, clearly distinguish between \spin0 and \spin2 resonances. This
allows one to separate the two cases and establish the correct spin nature
of the observed resonance. Furthermore, the azimuthal-angle difference
between the two photons differs if the structure of the contributing
operators is changed. This can provide a handle on how a possible \spin2
state is coupled to the gauge bosons.

Among  the processes with four leptons and two jets in the final state, namely \\
  $e^+ \, e^- \, \mu^+ \mu^- \, jj$, $\, \, e^+ \, e^- \, \nu_\mu \overline{\nu}_\mu \, jj$, $\, \, e^+ \, \nu_e \, \mu^- \, \overline{\nu}_\mu \, jj$, $ \, \, e^+ \, \nu_e \,
 \mu^+ \mu^- \, jj$ and $e^- \, \overline{\nu}_e \, \mu^+ \mu^- \, jj$ production,
\spin2 resonances can be observed most cleanly in the first of them, albeit at
a very low rate, since the resonance can be
reconstructed in the invariant-mass distribution of the four final-state charged
leptons. Heavy \spin2 resonances feature characteristic differential distributions, which are 
observable if the couplings are not too small and a proper cut on the invariant four-lepton mass is applied. The distribution of the transverse 
momentum of the hardest final-state lepton, the azimuthal angle difference between the two tagging jets, the cosine of the angle 
between the momenta of an incoming 
and an outgoing electroweak boson in the rest frame of the \spin2 resonance and the pseudorapidity difference between two final-state leptons are especially 
useful to identify a \spin2 resonance above the SM background. 
In the other processes with four leptons and two jets in the final
state, heavy \spin2 resonances are only visible in the transverse-mass spectrum for relatively 
large \spin2 couplings. However, some characteristics of the differential distributions also remain accessible in the case of smaller couplings if appropriate 
transverse-mass cuts are applied. In the processes $pp \rightarrow V V \, jj \rightarrow e^+ \, \nu_e \, \mu^+ \mu^- jj$ and \mbox{$pp \rightarrow V V \, jj \rightarrow 
e^- \, \overline{\nu}_e \, \mu^+ \mu^- \, jj$}, only charged resonances
are possible. Thus, they can be useful to distinguish between the \spin2 singlet and 
triplet scenarios.

Similar to the SM case, the NLO QCD corrections are small, leading to
slightly enhanced cross sections with $K$-factors of approximately 
1.05 and have no impact on the characteristics of the
differential distributions.

\FloatBarrier\section*{Acknowledgments}
\noindent
This research was supported in part by the Deutsche
Forschungsgemeinschaft via the Sonderforschungsbereich/Transregio
SFB/TR-9 ``Computational Particle Physics'' and the Initiative and
Networking Fund of the Helmholtz Association, contract HA-101(``Physics at
the Terascale''). J.F.\ acknowledges support by the
"Landesgraduiertenf\"orderung" of the State of Baden-W\"urttemberg. 

\appendix
\FloatBarrier\section{Decay widths of the Spin-2 singlet and triplet \newline particles \label{sec:decaywidths}}
\subsection{Singlet}

We define the total decay width of the \spin2 singlet particle $T$ as
\begin{equation}
\Gamma_{\text{total}}=\sum_j \Gamma_j \cdot \frac{1}{b}, \label{widthdef}
\end{equation}
where the sum runs over the decay channels resulting from the relevant vertices of \mbox{Sec. \ref{sec:spin2model}}. The free parameter $b$, 
which is set to 1 in the present analysis, is the fraction of these decays over all possible decays.\\ 
\\
The explicit results for the partial decay widths, $\Gamma_j$, are:\\
{\allowdisplaybreaks
\begin{align}
\Gamma_{W^+W^-} &= \left( \frac{24 f_2^2(m_T^4-3 m_T^2 m_W^2
+6 m_W^4)+40 f_2 f_5 g^2 v^2(m_T^2
-m_W^2)}{12 \Lambda^2} \right. \nonumber\\
&\quad \left. +\frac{f_5^2 g^4 v^4 
(m_T^4+12 m_T^2 m_W^2+56 m_W^4)}
{96 \Lambda^2 m_W^4} \right) \cdot
 \frac{\sqrt{(m_T^2/4-m_W^2)}}
{(40 \pi m_T^2)}\\[0.5 cm]
\Gamma_{ZZ} &= \left( [24 f_2^2 c_w^4 (m_T^4- 3 m_T^2
 m_Z^2+ 6 m_Z^4)+8 c_w^2 f_2 
(6 f_1 s_w^2 (m_T^4-3 m_T^2
 m_Z^2+6 m_Z^4) \right.\nonumber\\
&\quad +5 f_5 v^2 (g^2+g'^2)
 (m_T^2-m_Z^2))+24 f_1^2 s_w^4 
(m_T^4-3 m_T^2 m_Z^2+6 m_Z^4)\nonumber\\
&\quad +40 f_1 f_5 s_w^2 v^2 (g^2+g'^2) 
(m_T^2-m_Z^2)]
/(12 \Lambda^2)\nonumber\\
&\quad \left.
+\frac{f_5^2 v^4 (g^2+g'^2)^2 
(m_T^4+12 m_T^2 m_Z^2+56 m_Z^4)}
{96 \Lambda^2 m_Z^4} \right)
\cdot \frac{\sqrt{m_T^2/4- m_Z^2}}
{80 \pi m_T^2}\\[0.3 cm]
\Gamma_{\gamma \gamma} &= \frac{(f_1 c_w^2+f_2 s_w^2)^2 m_T^3}{80 \pi \Lambda^2}\\[0.3 cm]
\Gamma_{\gamma Z} &= \frac{c_w^2 s_w^2 (f_1-f_2)^2 (m_T^2-m_Z^2)^3
 (6 m_T^4+3 m_T^2 m_Z^2+m_Z^4
)}{240 \pi \Lambda^2 m_T^7},
\end{align}}
where $m_T$ denotes the mass of the \spin2 singlet particle.

\subsection{Triplet}

For the decay width of the neutral and the charged \spin2 triplet particles, the same definition (Eq. (\ref{widthdef})) is applied. The parameter $b$ 
can differ from the singlet case and can be different for the neutral and the charged particles, yet it is set to 1 in the present analysis for all cases. 
\\
The following equations present the results for $\Gamma_j$ for the
neutral \spin2 triplet particle: 

{\allowdisplaybreaks
\begin{align}
\Gamma_{W^+W^-}&= \frac{f_6^2 g^4 v^4 
(m_T^4+12 m_T^2 m_W^2+56 m_W^4)}
{384 \Lambda^2 m_W^4}
\cdot \frac{\sqrt{m_T^2/4-m_W^2}}
{40 \pi m_T^2}\\[0.3 cm]
\Gamma_{ZZ}&= \left(\left(768 f_7^2 c_w^2 s_w^2 m_Z^4
(m_T^4-3 m_T^2
m_Z^2+6 m_Z^4)+640 c_w f_6 f_7 m_Z^4
s_w v^2
 \right. \right.\nonumber\\
&\quad \left. (g^2+g'^2) (m_T^2- m_Z^2)
+f_6^2 v^4 (g^2+g'^2)^2
(m_T^4+12 m_T^2 m_Z^2+56 m_Z^4)\right)\nonumber\\
&\quad \left.
/(384 \Lambda^2 m_Z^4)\right)
\cdot \frac{ \sqrt{m_T^2/4- m_Z^2}}
{80 \pi m_T^2}\\[0.3 cm]
\Gamma_{\gamma \gamma} &= \frac{f_7^2 c_w^2 s_w^2 m_T^3 }
{ 80 \pi \Lambda^2}\\[0.3 cm]
\Gamma_{\gamma Z} &= \frac{ f_7^2(c_w^2-s_w^2)^2 
(m_T^2- m_Z^2)^3 
(6 m_T^4+3 m_T^2 m_Z^2+ m_Z^4
)}{960  \pi \Lambda^2 m_T^7}.
\end{align}}
Here, $m_T$ denotes the mass of the neutral \spin2 triplet particle.\\
\\
The partial decay widths of the charged \spin2 particles are:
{\allowdisplaybreaks

\begin{align}
\Gamma_{W \gamma} &= \frac{f_7^2 c_w^2 (m_T^2- m_W^2)^3
(6 m_T^4+3 m_T^2 m_W^2+ m_W^4
)}{960 \pi \Lambda^2 m_T^7}
\end{align}
\newpage
\begin{align}
\Gamma_{WZ} &= ((m_T^2 m_W^2 (m_T^2 m_Z^2
(13 f_6^2 g^2 v^4 (g^2+g'^2)+256 f_7^2
 m_W^2 m_Z^2 s_w^2) \nonumber\\
 &\quad+1/4 (m_T^2- m_W^2
+ m_Z^2)^2 (7 f_6^2 g^2 v^4 (g^2+g'^2)-
96 f_7^2 m_W^2 m_Z^2 s_w^2))\nonumber\\
 &\quad+1/4 (m_T^2+ m_W^2- m_Z^2)^2 (m_T^2 m_Z^2
 (7 f_6^2 g^2 v^4 (g^2+g'^2)-96 f_7^2 
 m_W^2 m_Z^2 s_w^2)\nonumber\\
 &\quad +(m_T^2- m_W^2+ m_Z^2)^2
(f_6^2 g^2 v^4 (g^2+g'^2)+32 f_7^2 m_W^2 
m_Z^2 s_w^2))\nonumber\\
 &\quad+m_T^2 (m_T^2- m_W^2-
 m_Z^2) (1/4 (m_T^2+ m_W^2- m_Z^2) 
(m_T^2- m_W^2+ m_Z^2)\nonumber\\ 
 &\quad(128 f_7^2 m_W^2 m_Z^2
 s_w^2-f_6^2 g^2 v^4 (g^2+g'^2))+40 f_6 f_7 g 
m_T^2 m_W^2 m_Z^2 s_w v^2 \sqrt{g^2+g'^2
})\nonumber\\
 &\quad+ 1/4 m_T^4 (-m_T^2+ m_W^2
+ m_Z^2)^2 (f_6^2 g^2 v^4 (g^2+g'^2)+32 f_7^2
 m_W^2 m_Z^2 s_w^2)\nonumber\\
 &\quad+40 f_6 f_7 g m_T^2 m_W^2
 m_Z^2 s_w v^2 \sqrt{g^2+g'^2} (m_T^2
+ m_W^2-m_Z^2)\nonumber\\
&\quad (m_T^2-m_W^2+ m_Z^2)) /(96 \Lambda^2 m_T^4 m_W^2 m_Z^2))\nonumber\\
&\quad \cdot \frac{\sqrt{(m_T^2- m_W^2- m_Z^2)^2- 4 m_W^2 m_Z^2
}}{80 \pi m_T^3},
\end{align}}
with $m_T$ now being the mass of the charged \spin2 triplet particles.


\end{document}